\PassOptionsToPackage{warn}{textcomp}
\documentclass[acmsmall,screen]{acmart}

\setcopyright{rightsretained}
\acmPrice{}
\acmDOI{10.1145/3341696}
\acmYear{2019}
\copyrightyear{2019}
\acmJournal{PACMPL}
\acmVolume{3}
\acmNumber{ICFP}
\acmArticle{92}
\acmMonth{8}

\usepackage[utf8]{inputenc}
\usepackage[T1]{fontenc}
\usepackage{microtype}

\usepackage[english]{babel} 
\usepackage{amsmath,amsthm} 
\usepackage{amssymb}

\usepackage{bm}
\usepackage{ stmaryrd } 

\usepackage[normalem]{ulem}
\usepackage{mathpartir}
\usepackage{color}

\newcommand{\THESYSTEM}{ARel}
\newcommand{\Relcost}{RelCost}
\makeatletter 
\def\arcr{\@arraycr}
\makeatother

\newcommand{\grt}{A}

\newcommand{\tbool}{\mbox{bool}}

\newcommand{\trint}{\mbox{int}}
\newcommand{\tint}{\mbox{int}}
\newcommand{\tunit}{\mbox{unit}}
\newcommand{\trunit}{\mbox{unit}}

\newcommand{\tarrd}[1]{\mathrel{\xrightarrow{\wdiff(#1)}}}

\newcommand{\tbox}[1]{\square\,#1}

\newcommand{\eapp}{\;}

\newcommand{\einl}{\mbox{inl\;}}
\newcommand{\einr}{\mbox{inr\;}}

\newcommand{\ecase}{\mbox{\;case\;}}

\newcommand{\eswitch}{\mbox{switch}\;}

\newcommand{\esplit}{\mbox{split}\;}

\newcommand{\ealloc}[2]{ \mathrel{ \mathsf{alloc}\, {#1} \, {#2} } }
\newcommand{\eallocB}[2]{ \mathrel{ \mathsf{alloc_b}\, {#1} \, {#2} } }
\newcommand{\eupdt}[3]{ \mathrel{ \mathsf{updt} \ {#1} \ {#2} \ {#3} }  }
\newcommand{\ereadx}[2] { \mathrel{ \mathsf{read} \ {#1} \ {#2} }  }
\newcommand{\eupdtB}[3]{ \mathrel{ \mathsf{updt_b} \ {#1} \ {#2} \ {#3} }  }
\newcommand{\ereadxB}[2] { \mathrel{ \mathsf{read_b} \ {#1} \ {#2} }  }

\newcommand{\ewith}{\;\mbox{with}\;}

\newcommand{\rdiff}{\ominus}

\newcommand{\ldiff}{\lesssim}

\newcommand{\wdiff}{\mbox{\scriptsize diff}}

\newcommand{\jiterm}[2]{\mathrel{\vdash {#1} :: #2}}
\newcommand{\jtype}[4]{\mathrel{\vdash_{#1}^{#2} {#3} : #4}}

\newcommand{\rname}[1]{\mbox{\small{#1}}}

\newcommand{\cref}{\mathsf{ref}}

\newcommand{\jtypediff}[4]{\mathrel{\vdash 
    {#2} \ominus {#3} \ldiff #1 : {#4}}}

\newcommand{\chdiff}[5]{\vdash{#1}\rdiff{#2}~{\downarrow}~#3,#4 \Rightarrow
{\color{red}#5}}

\newcommand{\infdiff}[6]{\vdash{#1}\rdiff{#2}~{\uparrow}~{\color{red}{#3}}\Rightarrow[{\color{red}#4}],{\color{red}#5},{\color{red}#6}}

\newcommand{\chexec}[5]{\vdash{#1}~{\downarrow}~#2, #3, #4  \Rightarrow
{{\color{red}#5}}}

\newcommand{\infexec}[6]{\vdash{#1}~{\uparrow}~{\color{red}{#2}}\Rightarrow[{\color{red}#3}], {\color{red}#4},{\color{red}#5},{\color{red}#6}}

\newcommand{\senv}{\Delta}
\newcommand{\uenv}{\Omega}
\newcommand{\renv}{\Gamma} 
\newcommand{\cenv}{\Phi} 
\newcommand{\lenv}{\Sigma}
\newcommand{\sep}{ \, | \, }

\newcommand{\monadR}[4]{\mathrel{ \overset{\mathrm{diff}(#4)}{\{ {#1} \} \ {#2} \ \{ {#3} \} }}}

\newcommand{\uarrow}[3]{ \mathrel{ \stackrel{\mathrm{exec} {#3}}{{#1} \longrightarrow#2}}}
\newcommand{\uforall}[4]{ \mathrel{ \stackrel{\mathrm{exec} {#4}}{\forall {#1} {::}#2 . \ #3}}}
\newcommand{\uexist}[3]{ \mathrel{ \exists {#1} :: {#2} . \ {#3}}}
\newcommand{\rarrow}[3]{ \mathrel{ \stackrel{\mathrm{diff} \, #3}{{#1} \longrightarrow {#2}}}}

\newcommand{\rforall}[4]{ \mathrel{ \stackrel{\mathrm{diff}\, {#4} }{\forall {#1}{::}{#2} . \ {#3}}}}
\newcommand{\rexists}[3]{ \mathrel{ \exists {#1} {::}{#2} . \ {#3}}}

\newcommand{\arr}[3]{ \mathrel{ \mathsf{Array}_{#1}[{#2}] \ {#3}} }
\newcommand{\arrR}[3]{ \mathrel{ \mathsf{Array}_{#1}[{#2}] \ {#3}} }
\newcommand{\lst}[2]{ \mathrel{ \mathsf{list}[{#1}] \ {#2}} }
\newcommand{\lstR}[3]{ \mathrel{ \mathsf{list}^{#1}[{#2}] \ {#3}} }
\newcommand{\abs}[2]{\mathrel { \lambda {#1} . {#2} } }
\newcommand{\app}[2]{\mathrel{ {#1} \, {#2} }}
\newcommand{\ret}[1] {\mathrel{ \mathsf{return} \, {#1} }}
\newcommand{\letx}[3]{  \mathrel{ \mathsf{let} \{ {#1} \} = {#2} \
\mathsf{in} \ {#3}  } }
\newcommand{\letm}[3]{  \mathrel{ \mathsf{let}\   {#1} = {#2} \ \mathsf{in} \ {#3}  } }
\newcommand{\packx}[1]{  \mathrel{ \mathsf{pack} \, {#1}} }
\newcommand{\unpackx}[3]{  \mathrel{ \mathsf{unpack} \,  {#1} \, \mathsf{as} \, {#2} \, \mathsf{in} \ {#3}  } }
\newcommand{\alloc}[2]{ \mathrel{ \mathsf{alloc}\, {#1} \, {#2} } }
\newcommand{\updt}[3]{ \mathrel{ \mathsf{updt} \ {#1} \ {#2} \ {#3} }  }
\newcommand{\readx}[2] { \mathrel{ \mathsf{read} \ {#1} \ {#2} }  }

\newcommand{\tfix}{\mathsf{Fix}}
\newcommand{\fix}[1]{\mathsf{fix} \, f(x). {#1}}

\newcommand{\monadu}[4]{\mathrel{ \overset{ \mathrel{\mathrm{exec}{#4 }}}{\{ {#1} \} \ #2 \ \{ {#3} \}} }}
\newcommand{\cmp}[4] {\mathrel{   \vdash  {#1} \ominus {#2} \ldiff {#4}  : {#3}  }}

\newcommand{\eval}[3]{\mathrel{ {#1} \Downarrow^{#3} {#2}   }}
\newcommand{\evalf}[3]{\mathrel{ {#1} \Downarrow^{#3}_{f} {#2}   }}

\newcommand{\heap}[1]{ ;  {#1}}

\newcommand{\wfa}[1]{\mathrel{\vdash {#1} \ \mathsf{wf}}}
\newcommand{\wf}[1]{\mathrel{\vdash {#1} \ \mathsf{wf}}}
\newcommand{\subtypeA}[2]{\mathrel{ \models {#1} \sqsubseteq {#2} } }
\newcommand{\subtype}[2]{\mathrel{   \models {#1} \sqsubseteq {#2} }   }
\newcommand{\subcost}[3]{\mathrel{   \models {#1} {#3} {#2} }   }

\newcommand{\emptyhp}{\mathsf{empty}}
\newcommand{\llb}[1]{ \llbracket {#1} \rrbracket }
\newcommand{\llu}[2]{ \llb{#1}_{#2}}
\newcommand{\llp}[2]{ \llparenthesis {#1} \rrparenthesis_{#2} }

\newcommand{\stepi}{k}

\newcommand{\trmo}[1]{|#1|}
\newcommand{\bctx}{\Delta; \psi_a; \Phi_a; \Gamma}
\newcommand{\tcho}[1]{U\,#1}

\newcommand{\ceq}[2]{#1\mathrel{\doteq}#2}
\newcommand{\cleq}[2]{#1 \mathop{\leq} #2}

\newcommand{\cand}[2]{#1 \wedge #2}
\newcommand{\cexists}[3]{\exists#1::#2.#3}

\newcommand{\cimpl}[2]{#1\rightarrow#2}

\newcommand{\freshCost}[1]{#1\in\text{fresh}(\scost)}

\newcommand{\scost}{\mathbb{R}}

\newcommand{\wfcs}[1]{\vdash   #1~\mathsf{wf}}

\newcommand{\rd}[1]{\textcolor[rgb]{0,0.,0}{#1}}

\usepackage{tikz}
\newcommand{\Arrow}[1]{%
\parbox{#1}{\tikz{\draw[->](0,0)--(#1,0);}}
}

\usepackage{booktabs}   

\begin{document}


\title{Relational Cost Analysis for Functional-Imperative Programs}

\author{Weihao Qu}
\affiliation{
\institution {University at Buffalo, SUNY, USA} 
}
\email{weihaoqu@buffalo.edu} 

\author{Marco Gaboardi}
\affiliation{
  \institution{University at Buffalo, SUNY, USA}           
}
\email{gaboardi@buffalo.edu}

\author{Deepak Garg}
\affiliation{
  \institution{MPI-SWS, Germany}           
}
\email{dg@mpi-sws.org}

\begin{abstract}

Relational cost analysis aims at formally establishing bounds on the
difference in the evaluation costs of two programs. As a particular
case, one can also use relational cost analysis to establish bounds on
the difference in the evaluation cost of the same program on two
different inputs.
One way to perform relational cost analysis is to use a relational
type-and-effect system that supports reasoning about relations between
two executions of two programs.

Building on this basic idea, we present a type-and-effect system, called {\THESYSTEM}, for
reasoning about the relative cost of array-manipulating, higher-order
functional-imperative programs. The key
ingredient of our approach is a new lightweight type refinement
discipline that we use to track relations (differences) between two
mutable arrays. This discipline combined with Hoare-style triples built into
the types allows us to express and establish precise relative costs of
several interesting programs which imperatively update their data. We
have implemented {\THESYSTEM} using ideas from bidirectional type checking.

\end{abstract}

\ccsdesc[500]{Theory of computation~Type structures}
\ccsdesc[500]{Theory of computation~Program verification}


\keywords{relational type systems, refinement types, type-and-effect systems}  

\maketitle

\section{Introduction}
\label{sec:intro}
Standard cost analysis aims at statically establishing an upper or a
lower bound on the evaluation cost of a program. The evaluation cost
is usually measured in abstract units, e.g., the number of reduction
steps in an operational semantics, the number of recursive calls made
by the program, the maximum number of abstract units of memory used
during the program's evaluation, etc. Cost analysis has been developed
using a variety of techniques such as type
systems~\citep{HoffmannAH12a,lago12,Danielsson08,Dallago17,DMLcost},
term rewriting and abstract
interpretation~\citep{ciaopp-sas03-journal-scp,SinnZV14,BrockschmidtEFFG14},
and Hoare
logics~\citep{CarbonneauxHZ15,Atkey10,ChargueraudP15}.

Relational cost analysis, the focus of this paper, is a more recently
developed problem that aims at statically establishing an upper bound
on the \emph{difference} in the evaluation costs of two related
programs or two runs of the same program with different
inputs~\citep{Cicek17,NDFH17,DBLP:journals/pacmpl/RadicekBG0Z18}. This
difference is called the \emph{relative cost} of the two programs or
runs. Relational cost analysis has many applications: It can show that
an optimized program is not slower than the original program on
stipulated inputs; in cryptography, it can show that an algorithm's
run time is independent of secret inputs and, hence, that there are no
leaks on the timing side-channel; in algorithmic analysis, it can help
understand the sensitivity of an algorithm's cost to input changes,
which can be useful for resource allocation.

There are two reasons for examining relational cost analysis as a
separate problem, as opposed to performing standard unary cost
analysis separately on the two programs and taking a difference of the
established costs. First, in many cases, relational cost analysis is
easier than unary cost analysis, since it can focus
only on the differences between the two programs. As a trivial
example, if $t$ is a complex closed program, it may be very difficult
to perform unary cost analysis on it, but it is obvious that the cost
of $t$ relative to itself is $0$. Second, in many cases, a direct
relational cost analysis may be more precise than the difference of
two unary analyses, since the relational analysis can exploit
relations between intermediate values in the programs that the unary
analyses cannot. As an example, the relative cost of two runs of merge
sort on lists of length $n$ that differ in at most $k$ positions is in
$O(n \cdot (1 + \log(k)))$. This relative cost can be established by a
direct relational analysis~\citep{Cicek17}, but two separate unary
analyses can only establish the coarser relative cost
$O(n \cdot \log(n))$.

Hitherto, literature on relational cost analysis has been limited to
functional languages. However, many practical programs are stateful
and use destructive updates, which are more difficult to reason
about.  Consequently, our goal in this work is to develop relational
cost analysis for functional languages with mutable state (i.e., for
functional-imperative programs).

To this end, we design a \emph{refinement} \emph{type-and-effect}
system, {\THESYSTEM}, for relational cost analysis in a functional,
higher-order language with mutable state. The first question we must
decide on is \emph{what} kind of state to consider. 
\rd{One option could be to work with standard references as found in many functional
languages like ML. However, from the perspective of cost analysis it is
often more interesting to consider programs that operate on entire \emph{data
structures} (e.g., a sorting algorithm), not just on individual
references. 
}
Consequently, we consider \emph{mutable arrays}, the
standard data structure available in almost all imperative
languages. This makes our type system more complicated than it would
be with standard references but allows us to verify more interesting
examples.

Second, we must decide \emph{how} to treat state in our functional
language. Broadly, we have two choices: State could be a pervasive
effect as in ML, or it could be confined to a monad as
in Haskell, which limits the side-effect to only those
sub-computations that actually access the heap. In {\THESYSTEM}, we
choose the latter option since this separates the pure and impure
(state-affecting) parts of the language at the level of types and
reduces the complexity of our typing rules.

The primary typing judgment of {\THESYSTEM},
$\jtypediff{r}{t_1}{t_2}{\tau}$,  states that the programs
$t_1$ and $t_2$ are related at type $\tau$, which can specify
relational properties of their results and, importantly, that their
relative cost (cost of $t_1$ minus the cost of $t_2$) is upper-bounded
by~$r$.\footnote{This judgment is inspired by~\citet{Cicek17} proposing a type-and-effect system for relational cost
analysis of functional programs \emph{without state}. \rd{Notice that
one can use this typing judgment also to reason about \emph{lower
bounds} on the relative cost, by exchanging $t_2$ and $t_1$ and
considering a negative cost $-r$.}}
To reason about array-manipulating programs, we also need to express
relations between corresponding arrays across the two runs. For this,
our monadic type (the type of impure computations that can access
state) has a \emph{refinement} that specifies how arrays are related
across the two runs \emph{before} and \emph{after} a heap-accessing
computation. Specifically, our monadic type has the form
$\monadR{P}{\exists \vec{\gamma}.\tau}{Q}{r}$. This type represents a pair of computations which,
when starting from arrays related by the relational pre-condition $P$,
end with arrays related by the relational post-condition $Q$, return
values related at $\tau$, newly
generated arrays referred by static names $\vec{\gamma}$, and have relative cost at most
$r$. This design is inspired by relational Hoare
logics~\citep{rhtt,Benton04}, but there are two key differences:
1)~Our pre- and post-conditions are \emph{minimal}---they only specify the
indices at which a pair of arrays differ across the two runs, not full
functional properties. This suffices for relational cost analysis of
many programs and simplifies our metatheory and, importantly, the implementation.  2)~Our monadic types
carry a relative cost, and the monad's constructs combine and
propagate the different costs.



Additionally, {\THESYSTEM} supports establishing lower and upper
bounds on the cost of a \emph{single} expression, and falling back to
such unary analysis in the middle of a proof of relative
cost. Improving over previous type-and-effect systems for
relational cost analysis, {\THESYSTEM}  permits combinations of these two
kinds of reasoning in the definition of recursive
functions. {\THESYSTEM} provides typing rules for the fix-point operator that
allow one to \emph{simultaneously} reason about unary and relational cost. This is
useful for the analysis of several programs.

To prove that our type system is sound, we develop a logical relations
model of our types. This model combines unary and binary logical
relations and it supports two different effects, cost and state, that
are structurally dissimilar. For the state aspect, we build on
step-indexed Kripke logical
relations~\citep{DBLP:conf/popl/AhmedDR09, Ahmed2004SemanticsOT}. Specifically, our logical
relations are indexed by a ``step''---a standard device for inductive proofs
that counts how many steps of computation the logical relation is good
for~\citep{ahmed06:sll,AppelM01:si}. Owing to the simplicity of our
pre- and post-conditions, we do not need state-dependent worlds as in
some other work~\citep{DBLP:journals/jfp/NeisDR11,DBLP:conf/popl/TuronTABD13}.

To show the effectiveness of our approach, we implemented a
bidirectional type-checker for {\THESYSTEM}. Thanks to the simplified
form of our pre- and post-conditions, we can solve the constraints
generated by the type-checker using SMT solvers. The type-checker also
uses a restricted number of heuristics in order to address some of
the non-determinism coming from the relational reasoning. 
We used our implementation to type-check a broad set of examples
showing some of the challenges of relational cost analysis in programs
manipulating arrays.


\medskip

Our overarching contribution lies in extending relational cost
analysis to higher-order functional-imperative programs. Our specific
contributions are:
\begin{itemize}
\item {\THESYSTEM}, a  type system for relational cost
  analysis of functional-imperative programs with mutable arrays.
\item A design for lightweight (relational) refinements of
  array-based computations.
\item A soundness proof for our type system via a new step-indexed logical
  relation.
\item An implementation of {\THESYSTEM}, based on bidirectional type
  checking, which we use to type check several
functional-imperative examples.
\end{itemize}

\section{\THESYSTEM\ through examples}
\label{sec:overview}
In this section, we illustrate the key ideas behind {\THESYSTEM}
through two simple examples.

\paragraph*{Inplace Map} Consider the following imperative $\mathsf{map}$ function taking
as input a pure function $f$, a mutable array $a$, an index $k$ and
the array's length $n$. For all $i \in [k,n]$, the function replaces
the current value in the $i$th cell of $a$ with $f(a[i])$, thus
performing a destructive update.
%
\[
\begin{array}{l@{}l}
 \mathsf{fix} \ \mathsf{map} \,  ( f).  \lambda  a . \lambda k. \lambda n. \; &
 \mathsf{if} \ k \leq n  \ \mathsf{then} \
  \big ( \mathsf{let} \ \{x\} = \; \readx{a}{k}  \, \mathsf{in} \\ 
&   \hspace{.6cm} \mathsf{let} \{ \_ \} =\; \updt{a}{k}{(f \ x)}  \ \mathsf{in} \
   \mathsf{map} \, f \, a \,  (k+1) \, n \big)\,\\
& \mathsf{else} \  \mathsf{return} ()  
\end{array}
\]
The expression $(\readx{a}{k})$ returns the element at index $k$ in
the array $a$, and $(\updt{a}{k}{v})$ updates the index $k$ in $a$ to
$v$. Our language uses a state monad to isolate all side-effects like
array reads and updates, so $(\readx{a}{k})$ and $(\updt{a}{k}{v})$ are
actually expressions of monadic types, also called
\emph{computations}. The construct
$(\mathsf{let}\ \{x\} = t_1 \, \mathsf{in}\, t_2)$ is monadic
sequencing, often called ``bind''.

Consider the problem of establishing an upper-bound on the
\emph{relative} cost of two runs of $\mathsf{map}$ that use the
\emph{same} function $f$ but two \emph{different} arrays
$a$. Intuitively, the relative cost should be upper-bounded by the
product of the maximum variation in the cost of the function
$f$ (across inputs) and the number of indices in the range $[k,n]$ at
which the two arrays differ.

To support reasoning about two runs as in this example, {\THESYSTEM}
supports \emph{relational types} that ascribe a pair of related values
or related expressions in the two runs. Relational types are written
$\tau$. In general, when we say $x: \tau$, we mean that the variable
$x$ may be bound to two different values in the two runs, but these
two values will be related by the type $\tau$. 
\rd{Specifically, $x:
\tau_1 \to \tau_2$ means that $x$ can be bound to two different
functions $f_1, f_2$ in the two runs, satisfying the property that for
any two $v_1, v_2$ of relational type $\tau_1$, the two expressions $f_1\ v_1, f_2\ v_2$
have relational type $\tau_2$.}  
{\THESYSTEM} also supports \emph{unary
  types}, denoted $A$, that ascribe a value or expression in a single run,
\rd{but
we will have no occasion to use unary types in this example, so we
postpone their discussion.}

To establish the relative cost of $\mathsf{map}$, we first need a way
to represent that the \emph{same} function $f$ will be given to
$\mathsf{map}$ in both runs. For this, {\THESYSTEM} offers the type
annotation $\square$. The type
$\square \tau$ relates expressions in two runs that are
(syntactically) equal and are additionally related at the relational
type $\tau$. Note
that $\square$ is a \emph{relational refinement}: It refines the
relation defined by the underlying type $\tau$.
Specifically, the relational typing assumption $f:\square(\tau_1 \to \tau_2)$ means that, in the two runs, $f$ will be bound to two copies of the \emph{same} function, say $f$, that given arguments
$v_1, v_2$ related at type $\tau_1$, give expressions $f \ v_1$ and
$f \ v_2$ related at type $\tau_2$. In our example, if the array's elements have type $\tau$, the type of $f$ would be $\square (\tau \to \tau)$. 

Next, we need to represent the maximum possible variation in the cost
of \rd{applying} $f$. \rd{The possible variation in the cost can be seen
  as an \emph{effect}, and the cost of applying a function can be seen
  as the effect associated with the body of the function, in
  particular. Hence as is common in effect systems~\cite{nielsen99}, we can record the
  possible variation in cost by means of a refinement of the function
  type. {\THESYSTEM} offers a refinement of this kind.} 
We write $\rarrow{\tau_1}{\tau_2}{(r)}$ to represent
two functions of \rd{relational} type $\tau_1 \to \tau_2$, the relative
cost of whose bodies is upper-bounded by $r$. Accordingly, if $f$'s
cost can vary by $r$, its type can be further refined to
$\square (\rarrow{\tau}{\tau}{(r)})$.

Next, we need a way to specify \emph{where} the arrays given as inputs
to $\mathsf{map}$ in the two runs differ. There are various design
choices for supporting this. One obvious but problematic option would
be to refine the type of an array itself, to specify where the two
ascribed arrays differ across two runs. However, this design quickly
runs into an issue: An update on the arrays might be different in the
two runs, so it might change the arrays' \emph{type}. This would be
highly unsatisfactory since we don't expect the type of an array to
change due to an update; in particular, this design would not satisfy
(semantic or syntactic) type preservation.

Consequently, we use a different approach inspired by relational Hoare
logics: We provide a relational refinement type
$\monadR{P}{\exists\vec{\gamma}.\tau}{Q}{r}$ for monadic
expressions that manipulate  
arrays. The number $r$ is an upper-bound on the
relative cost of the computation, similar to the one we
have in function types,  and $\tau$ is the relational type of
the pure values the computation returns.
 The \emph{pre-condition} $P$ specifies for each pair of
related arrays in scope where (at which indices) the arrays are
allowed to differ \emph{before} the computation
runs, while the \emph{post-condition} $Q$ specifies where the arrays
may differ \emph{after} the computation completes. More specifically,
$P$ and $Q$ are lists of annotations of the form $\gamma \rightarrow
\beta$, where $\gamma$ is a \emph{static name} for an array and
$\beta$ is a set of indices where the array identified by
$\gamma$ \emph{may} differ in the two runs. At any index not in
$\beta$, the array must be the same in the two runs. Note that even at
indices in $\beta$, the corresponding values must be related at
$\tau$, but our type system includes types that do not force equality
of the related values. One such type is $U(A, B)$ that only insists
that the left and right values have (unary) types $A$ and $B$, without
requiring any other relation between them. (The existentially
quantified $\vec{\gamma}$ in
$\monadR{P}{\exists\vec{\gamma}.\tau}{Q}{r}$ is the list of static
names of arrays that are allocated during the computation.)

For example, if $x: \square \tau$, i.e., $x$ is the same in two runs,
and $b$ is an array of static name $\gamma$, then $(\updt{b}{5}{x})$
can be given the type
$\monadR{\gamma \rightarrow \beta}{\exists \_.\tunit}{\gamma \rightarrow
(\beta \setminus \{5\})}{0}$ relative to itself for any
$\beta$.\footnote{As usual, $\_$ represents a variable whose name is
unimportant.} This type means that if the array $b$ differs at the set
of indices $\beta$ before $(\updt{b}{5}{x})$ executes in two runs,
then afterwards it can still differ in the indices
$\beta$ \emph{except} at the index $5$, which has been overwritten by
the same value $x$. If we replace the assumption $x:
\square \tau$ with $x: \tau$, so that $x$ may differ in the two runs,
then the type of $(\updt{b}{5}{x})$ relative to itself would be
$\monadR{\gamma \rightarrow \beta}{\exists \_.\tunit}{\gamma \rightarrow (\beta
  \cup \{5\})}{0}$, indicating that the arrays may differ at index $5$
after the update (even if they did not differ at that index before the
update).

We also need a way to tie static names $\gamma$ appearing in
computation types to specific arrays. For this, we refine the type of
arrays to include $\gamma$. In fact, we also refine the type of arrays
to track the length of the array. This doubly refined type is written
$\arr{\gamma}{l}{\tau}$---a pair of arrays of length $l$ each,
identified statically by the name $\gamma$, and carrying elements
related pointwise at type $\tau$. Finally, we refine integers very
precisely: The type $\tint[n]$ is the \emph{singleton type} containing
only the integer $n$ in both runs. The $n$ in the type is a static
representation of the runtime values the type ascribes.

With all these components we can now represent 
the relative cost of
$\mathsf{map}$ that we are interested in by the
judgment:
\begin{multline*}
{  \jtypediff{0}{\mathsf{map}}{ \mathsf{map}}{} \begin{array}{l} \forall
  r:.\square (\rarrow{\tau}{\tau}{(r)}) \rightarrow \forall k,n,
  \gamma, \beta. (k\leq n) \supset \\ \arr{\gamma}{n}{\tau} \rightarrow
  \tint[k] \rightarrow \tint[n]  \rightarrow {\monadR{\gamma
      \rightarrow \beta}{\exists \_ .\tunit}{\gamma \rightarrow \beta}{|\beta \cap
      [k,n]|*r}}
\end{array}}
\end{multline*}

This typing means that $\mathsf{map}$ relates to itself in the
following way. Consider two runs of $\mathsf{map}$ with the same
function $f$ of relative cost $r$ (type $\square
(\rarrow{\tau}{\tau}{(r)})$), two arrays of static length $n$,
statically named $\gamma$ (type $\arr{\gamma}{n}{\tau}$), two indices,
both $k$ (type $\tint[k]$), and two lengths, both $n$ (type
$\tint[n]$). Then, the two runs return computations with the following
relational property: If the two arrays differ at most at indices
$\beta$ before they are passed to $\mathsf{map}$, then they differ at
most at the same positions after the computations and the relative
cost of the two computations is upper-bounded by $|\beta \cap
[k,n]|*r$, i.e., the number of positions in the range $[k,n]$ at which
the arrays may differ times $r$. This is exactly the expected relative
cost because at positions where the arrays are equal, $f$ will have
the same cost in the two runs (we are assuming language-level
determinism here).  Note that the variables $r, k, n, \gamma$ and
$\beta$ are universally quantified in the type above. Also note how
$\gamma$ links the input array to the $\beta$ in the pre- and
post-condition of the computation type.


Consider now a slightly different situation where
\emph{different} functions $f$ may be passed to $\mathsf{map}$ in the
two runs. Suppose that the relative cost of the bodies of the two $f$s
passed is upper-bounded by $r$, i.e., $f$ has the type
$\rarrow{\tau}{\tau}{(r)}$ (without the prefix $\square$). In this case,
the relative cost of the two runs of $\mathsf{map}$ can only be
upper-bounded by $|[k,n]| * r$, since even at indices where the arrays
agree, the cost of applying the two different $f$s may differ by as
much as $r$. Moreover, the final arrays may differ in all positions in
the range $[k,n]$. This is formalized in the following, second
relational type for $\mathsf{map}$.
\begin{multline*}
{  \jtypediff{0}{\mathsf{map}}{ \mathsf{map}}{} \begin{array}{l} \forall r. (\rarrow{\tau}{\tau}{(r)}) \rightarrow \forall k,n,
\gamma, \beta. (k\leq n)\, \supset \\  \arr{\gamma}{n}{\tau}
\rightarrow \tint[k] 
 \rightarrow \tint[n]  \rightarrow {\monadR{\gamma \rightarrow
                                                  \beta}{\exists \_ .\tunit}{\gamma \rightarrow \beta \cup [k,n]}{(n-k)*r}}
\end{array}}
\end{multline*}

\paragraph*{Boolean Or} 
Next, we describe how high-level reasoning about
relative cost is internalized in the typing. 
{\THESYSTEM} supports two kinds of typing modes: \emph{relational
  typing} as shown in the map example above, and \emph{unary
  typing} which supports traditional (unary) min- and
max-cost analysis for a single run of a program. 
We will introduce these modes formally in the next
section but here we want to show with the following example how they can be meaningfully combined.
%
\[
\begin{array}{@{}l@{}l@{}}
 \textsf{fix} \, \mathsf{BoolOr} \, (a). \, \lambda k. \lambda n. &
  \mathsf{if} \, k < n \, \mathsf{then}\, \big (\mathsf{let}  \{ x \} =
  \readx{a}{k} \mathsf{in}\ 
  \mathsf{if} \, x \, \mathsf{then}\, \mathsf{return} \,
  true \, \mathsf{else}  \, \mathsf{BoolOr} \, a  \, (k+1)   n\big)\\
 & \mathsf{else} \,  \mathsf{return} \, false
\end{array}
\]
This function, given as input an array of booleans $a$, an index $k$ and the array's
length $n$ tells whether there
exists an element in $a$ with index $\geq k$ and value \emph{true}. 


 Given two arbitrary arrays $a$ in two runs, a simple upper-bound
on the relative cost of $\mathsf{BoolOr}$ is $(n-k)*c$ where $c$ is
the cost of one iteration. This is because in one run we can find an element with value
$true$ in position $k$, and so the computation can return immediately,
while in the other run we may not find any such element, and would
need to visit every element of the array with its index greater than $k$.
This kind of high-level reasoning corresponds to a worst-case, best-case analysis of
the two individual runs.  
{\THESYSTEM} supports this kind of reasoning by supporting worst-case,
best-case (unary) cost analysis in unary mode, and 
by means of a rule \rname{R-S}, presented formally in Section~\ref{sec:syntax},
allows us to derive a relational typing from two unary typings, with
relative cost equal to the difference between the max and the min costs of the
unary typings. 

However, this kind of reasoning does not account for the case where
the two input arrays have a meaningful relation, e.g., they may be equal in some positions. In such cases, a
better upper bound on the relative cost would be expressed in term of
the first index $i$ (if any) where the two arrays differ. That is, we could
have the upper bound $(n-i)*c$. 
Showing this upper bound in a formal way is more
involved. We first need to proceed by case analysis on whether the element
$x$ we are reading at each step is the same in the two runs or not. 
Case analysis in {\THESYSTEM} is provided by the rule \rname{R-P}, presented in
Section~\ref{sec:syntax}. Using this rule we can consider the two
cases separately in typing the subexpression
$\mathsf{if} \ x \ \mathsf{then}\, (\mathsf{return} \,
  true) \, \mathsf{else}  \, \mathsf{BoolOr} \, a  \, (k+1)  \,
  n$.

If $x$ is the same in the two runs, there is no difference in cost
because we either $\mathsf{return} \, true$ in both runs or we perform
the recursive call in both runs. In case the two $x$'s differ, we must
switch to unary analysis of the two individual runs, since in one run
we will return immediately while in the other we will make a recursive
call, so there is no way to continue reasoning relationally. Hence, in
order to derive the required upper bound on the overall relative cost
we need to have information about the \emph{unary} type of
$\mathsf{BoolOr}$. However, since we started by trying to type the
body of $\mathsf{BoolOr}$ relationally, the standard fixpoint rule
only allows us to assume its \emph{relational} type.

One solution to this impasse is to automatically transform relational
types of variables in context to unary types when switching from
relational to unary reasoning. This approach was adopted
by~\citet{Cicek17} for analyzing pure functional programs but it
provides only trivial lower and upper bounds ($0$ and $\infty$) on the
costs of function variables in the context during the unary analysis.
In our example here, this approach yields the trivial upper bound
$\infty$, which is not what we want.

To allow for a more precise analysis, {\THESYSTEM} includes a new
rule \rname{R-FIX-EXT} which we introduce formally in
Section~\ref{sec:syntax}. This rule allows us to assume the result of
a \emph{unary} typing of two recursive functions, when typing their
bodies \emph{relationally}. With this rule, we can use the (assumed)
relational type of $\mathsf{BoolOr}$ and its unary type in typing the
subexpression $\mathsf{BoolOr} \, a \, (k+1) \, n$.  Hence, we can
conclude the inductive step and assign the precise relative cost
$(n-i)*c$ to $\mathsf{BoolOr}$.

\begin{figure*}
$
\begin{array}{l}
\begin{array}{rl}
    \text{\bf Index terms} & I, L, U,D, \alpha,\beta ::=  i
                                    \sep b\sep n \sep r\sep 
                                     I_{1}+I_{2} 
\sep I_{1} *I_{2} \sep I_{1} - I_{2} 
\sep max(I_{1}, I_{2}) \sep min(I_{1}, I_{2}) \\ & 
\sep log_{2}(I)  \sep \lfloor I \rfloor \sep \lceil I \rceil\sep
   \{ I_i \}_{i \in K} \sep \beta \cup \beta \sep \beta \setminus \beta \sep \beta \cap \beta
  \\[1mm]
    \text{\bf Terms} & t ::=  x\sep n \sep r \sep () \sep  
                                   \abs{x}{t} \sep \fix{t} \sep  \app{t_1}{t_2} \sep
                       \letm{x}{t_1}{t_2} \sep \einl t \sep \einr  t
  \\ &   
  \sep \ecase(t, x.t_{1}, y.t_{2}) \sep \Lambda.t \sep t\, [] \sep \packx{t} \sep
                       \unpackx{t_{1}}{x}{t_{2}} \sep
                       \mathsf{celim} \, t 
\\ & 
\sep  \ret{t} \sep \letx{x}{t_1}{t_2}  
   \sep \alloc{t_1}{t_2} \sep \readx{t_1}{t_2} \sep
                                   \updt{t_1}{t_2}{t_3}  \\[1mm]
    \text{\bf Values} & v ::=  n \sep l \sep r  \sep () \sep
                                     \abs{x}{t} \sep \fix{t} \sep
                        \einl v \sep \einr v \sep \Lambda . t \sep
                        \packx{v}  \\& \sep \ret{t}  \sep
                                     \alloc{t_1}{t_2} \sep 
                               \updt{t_1}{t_2}{t_3} \sep
                                                               \readx{t_1}{t_2} \sep \letx{x}{t_1}{t_2} \\[1mm]
  \text{\bf Unary types} & A ::= c \sep \tint[I]\sep 
          \monadu{P}{\exists \vec{\gamma}. A}{Q}{(L,U)} \sep \uforall{i}{S}{A}{(L,U)}\sep  \uexist{i}{S}{A}\sep
                             \uarrow{A}{{A}}{(L,U)} \sep
                           \arr{\gamma}{I}{A} \\& \sep
          \lst{I}{A} \sep
A_{1} + A_{2} \sep C \& A \sep C \supset A  \\[1mm]
  \text{\bf Relat. types} & \tau ::= c \sep
                                              \trint[I] |
                                             \monadR{P}{\exists \vec{\gamma}.\tau}{Q}{D}
                                             | \rforall{i}{S}{\tau}{(D)} \sep  \rexists{i}{S}{\tau} \sep
                            \rarrow{\tau}{{\tau}}{(D)} \sep
                            \arrR{\gamma}{I}{\tau}  \\& \sep
                                                        \lstR{\alpha}{I}{\tau}
                                                        \sep  \tau_{1}
                                                        + \tau_{2}
                                                        \sep  C \&
                                                                   \tau  \sep  C \supset \tau \sep U(A_1,A_2) \sep 
\square \tau  \\[1mm]
\end{array}\\
\begin{array}{rcl}
    \text{\bf Unary Type Env.} &\quad &  \uenv ::= \emptyset \sep  \uenv, x:A \\[1mm]
    \text{\bf Relational Type Env.} &\quad &  \renv ::= \emptyset \sep  \renv, x:\tau \\[1mm]
  \text{\bf Sort Env.} &\quad & \senv ::= \emptyset \sep \senv, i::S  \\[1mm]
  \text{\bf Loc Env.} &\quad & \lenv ::= \emptyset \sep \lenv, \gamma :: \mathbb{L} \\[1mm]
    \text{\bf Sorts} && S ::= \mathbb{R}\sep \mathbb{N}\sep
                        \mathbb{B}\sep \mathbb{P} \sep \mathbb{L} \\[1mm]
    \text{\bf Constraints } &\quad & C ::= I_{1} = I_{2} \sep I_{1} <
                                     I_{2} \sep \neg C \sep I_1\in I_2 \\[1mm]
    \text{\bf Constraint Env.} &\quad & \cenv ::= \top \sep C \wedge \cenv \\[1mm] 
   \text{\bf Assertions} & \quad & P , Q ::= \emptyhp \sep
                                   \gamma\rightarrow \beta \sep P \star Q\\[1mm]
    \text{\bf Heaps  } & \quad & H ::= [] \sep [ l \rightarrow z] | H_1 \uplus H_2 \\[1mm]
    \text{\bf Arrays } & \quad & z ::= [v_1,\ldots,v_m] \\[1mm]
\end{array}
\begin{array}{rl}
      & \text{ \bf Judgments } \\[1mm]
     &   \lenv; \senv ; \cenv; \uenv \jtype{L}{U}{t}{A} \\[1mm]
    &    \lenv; \senv ; \cenv; \renv \jtypediff{D}{t_1}{t_2}{\tau}\\[1mm]
    &   \lenv; \senv \jiterm{I}{S} \\[1mm]
    &   \lenv;\senv;\cenv \wfa{A} \\[1mm]
    &   \lenv;\senv;\cenv \wf{\tau} \\[1mm]
    &   \senv \wf{C} \\[1mm]
     &   \uenv \wf{H} \\[1mm]
     &   \lenv ; \senv \wf{ P} \\[1mm]
    &    \lenv;\senv ; \cenv \subtypeA{A_{1} }{ A_{2} } \\[1mm]
    &   \lenv;\senv ; \cenv \subtype{ \tau_{1} }{ \tau_{2} }   \\[1mm]
\end{array}
\end{array}
\vspace{-2mm}
$
\caption{Syntax of \THESYSTEM\ where $n \in \mathbb{N},\ r\in
  \mathbb{R},\ x \in Var ,\ i \in iVar ,\ \gamma \in iLoc, \ l\in Loc$.}
\label{fig:syntax}
\end{figure*}

\begin{figure*}
\begin{mathpar}
  \inferrule*[ right=e-v]
  { }
  { \eval{ v   }{ v  }{0,0}}   
\and
  \inferrule*[ right=e-a]
  {
    {\eval{ t_1 }{ \abs{x}{t'} }{c_{1},\stepi_1 } }
    \quad
    { \eval{ t_2  }{ v    }{c_{2},\stepi_2 }}
    \quad
    {\eval{ t'[v/x]  }{ v_1 }{c_{3}, \stepi_3}  }
}
  { \eval{ \app{t_1}{t_2}  }{ v_1  } { c_{1} +
      c_{2}+c_3+c_{\mathrm{app}} , \stepi_1 + \stepi_2 +\stepi_3 +1 }    }
\vspace{-3mm}
\\
  \inferrule*[ right=e-f]
  {
    {\eval{ t_1 }{ \mathsf{fix}\, f\, x.t' }{c_{1}, \stepi_1 }}
    \quad
    { \eval{ t_2  }{ v    }{c_{2}, \stepi_2}}
    \quad
    {\eval{ t'[\mathsf{fix}\, f\, x.t'/f][v/x]  }{ v_1 }{c_{3}, \stepi_3} }
}
  { \eval{ \app{t_1}{t_2}  }{ v_1  } { c_{1} +
      c_{2}+c_3+c_{\mathrm{fapp}}, \stepi_1 + \stepi_2 +\stepi_3 +1}    }
\vspace{-3mm}
\\
  \inferrule*[ right=f-e]
  { 
    {\eval{ t_{1} }{ v }{ c_{1}, \stepi_1}  }
    \quad
    { \evalf{ v \heap{H}  }{  v_1 \heap{H_{1}}   }{c_{2}, \stepi_2}   }
    \quad
    { \eval{  t_{2}[v_1 / x]  }{ v_2 }{c_{3}, \stepi_3}  }
    \quad
    {\evalf{ v_2 \heap{H_{1}}  }{ v_{3} \heap{H_{2}}  }{c_{4}, \stepi_4}  }
}
  { \evalf{ \letx{x}{t_{1}}{t_{2}}  \heap{H} }{ v_{3}  \heap{ H_{2} }
    }{ c_{1} + c_{2} + c_{3} +c_{4}+c_{\mathrm{let}}, \stepi_1 +
      \stepi_2 +\stepi_3 + \stepi_4 +1}  }
\vspace{-3mm}
\\
  \inferrule*[ right=f-u]
  {
    {\eval{ t_1}{ l }{c_1,\stepi_1}}
    \quad
    { \eval{t_2}{n}{c_2, \stepi_2}}
    \quad
    { \eval{t_3}{v}{c_3, \stepi_3}  }
}
  { \evalf{ \updt{t_1}{t_2}{t_3}\heap{H}}{()
            \heap{H(l)[n]\leftarrow v} }{ c_1 + c_2 + c_3
            +c_{\mathrm{update}}, \stepi_1 + \stepi_2 +\stepi_3 +1} }
\vspace{-4mm}
\\
  \inferrule*[ right=f-t]
  {  \eval{ t   }{ v    }{c,\stepi}  }
  { \evalf{ \ret{t} \heap{H}  }{ v \heap{H} }{ c+c_{\mathrm{ret}},
      \stepi +1}    }
\and
  \inferrule*[ right=f-r]
  {
   {\eval{ t_1 }{ l }{c_{1}, \stepi_1} }
    \quad
    { \eval{ t_{2}  }{  n_{2} }{ c_{2}, \stepi_2 }  }
    \quad
    {H(l)[n]=v } 
}
  { \evalf{ \readx{t_{1}}{t_{2}}  \heap{H} }{ v \heap{H } }{c_{1} +
      c_{2} + c_{\mathrm{read}} , \stepi_1 + \stepi_2 +1 }  }   
\vspace{-5mm}
\\
  \inferrule*[ right=f-l]
  {
    {\eval{ t_1 }{ n }{ c_{1}, \stepi_1} }
    \quad
    { \eval{ t_2 }{ v  }{c_{2}, \stepi_2}  }
\quad
z=[\overbrace{v,\ldots ,v}^{n}]\quad l\ \text{fresh}
}
  { \evalf{ \alloc{t_1}{t_2}  \heap{H} }{ l \heap{H\uplus [l\to
              z]} }{c_{1} + c_{2} + c_{\mathrm{alloc}}, \stepi_1
            +\stepi_2 +1 }  }   
\vspace{-2mm}
\end{mathpar}
\caption{Selection of rules for pure evaluation $\eval{t}{v}{c,\stepi}$, and forcing
  evaluation $\evalf{t\heap{H}}{v\heap{H'}}{c,\stepi}$.}
\label{fig:evaluation}
\end{figure*}

\section{\THESYSTEM\ formally}
\label{sec:syntax}

 \subsection{Syntax}
We summarize \THESYSTEM's syntax in Figure~\ref{fig:syntax}. 
The term language underlying \THESYSTEM\ is a simply typed
$\lambda$-calculus with recursion and constructs for mutable
arrays. Most of these constructs are inherited
from \Relcost~\cite{Cicek17}, a type system for relative cost analysis
in \emph{pure} functional programs. Following that work, \THESYSTEM\
also has type refinements in the style of
DML~\cite{xi99:dependent}. The term constructs $\Lambda.t$ and $ t\,
[]$, $ \packx{t} $ and $ \unpackx{t_{1}}{x}{t_{2}} $ correspond to the
introduction and elimination of universal and existential types. The
construct $ \mathsf{celim} \, t$ eliminates the constraint implication
$C \supset \tau$.

New here are the constructs to deal with arrays: for \emph{allocating}
arrays ($\alloc{t_1}{t_2}$, where $t_1$ specifies the number of array
cells to be allocated, and $t_2$ the initial value to be stored in
each array cell), for \emph{reading} from arrays ($\readx{t_1}{t_2}$,
where $t_1$ specifies the array to read from, and $t_2$ the position
in the array to read from), and for \emph{updating} arrays
($\updt{t_1}{t_2}{t_3}$, where $t_1$ specifies the array to be
updated, $t_2$ the position in the array to be updated, and $t_3$ the
value to be used for the update). All imperative (array-manipulating)
constructs are confined to a monad. The constructs $\ret{t}$ and
$\letx{x}{t_1}{t_2}$ are the usual return and bind of the monad.
Impure expressions are distinguished by monadic types, but not
syntactically distinguished in the syntax of expressions. Impure
expressions (expressions of monadic types) are values, but can
be \emph{forced} using a special forcing semantics that we describe
below. Finally, arrays are referenced through locations, $l \in
Loc$. Although locations do not appear in programs, they do show up
during evaluation, so they are included in the syntax.

\subsection{Operational Semantics}
We define a cost-annotated, big-step operational semantics for our
language. Part of this semantics is based on heap manipulation. We
represent heaps as mappings $H=[l_1\to z_1,\ldots,l_n\to z_n]$ from
memory locations to concrete arrays $z=[v_1,\ldots,v_n]$. The notation
$H(l)[n]=v$ expresses that the value $v$ is stored in the heap $H$ in
the array pointed by the location pointer $l$ at the index $n$, the
notation $H(l)[n]\leftarrow v$ represents the heap $H$ where the array
pointed by $l$ is updated with the value $v$ at index $n$, and the
notation $H_1\uplus H_2$, in the spirit of separation logic, denotes a
disjoint union of the heaps $H_1$ and $H_2$.
We give a selection of the evaluation rules in
Figure~\ref{fig:evaluation}. We have two kinds of 
evaluation judgments: \emph{pure evaluation} 
$\eval{t}{v}{c,k}$ states that the (pure) expression $t$ evaluates to
the value $v$ with cost $c$, using  $k$ steps, while
\emph{forcing evaluation} 
$\evalf{t\heap{H}}{v\heap{H'}}{c,k}$ states that the impure expression
$t$ can be forced in the heap $H$ to the value $v$ and to the updated
heap $H'$ with cost $c$, consuming $k$ steps.

Steps $k$ are a proof artifact, needed only in our soundness proof
that relies on a step-indexed logical
relation (Section~\ref{sec:lr}). We count a unit step for
every elimination and monadic construct. Readers may ignore steps for
now. The costs $c$ are what we seek to upper bound (relatively) using
our type system and are, therefore, important. At every elimination
form or monadic construct, the semantics add a construct-dependent
cost. For example, the cost $c_{\mathsf{app}}$ appearing in the rules
is the cost of an application. By changing these costs and setting
some of them to $0$, we can get different cost models. Our type system is parametric in the costs of individual
constructs.

%

Most of the pure evaluation rules are standard.
The forcing evaluation rules are used to evaluate impure (monadic)
expressions manipulating heaps (arrays). The rule \rname{F-T} forces
the evaluation of an expression $\ret{t}$ by evaluating the underlying
pure expression $t$ using the pure evaluation semantics. The cost
consists of the cost of the pure evaluation of $t$ and the constant
cost $c_{\mathsf{ret}}$ for the monadic return.  The rule \rname{F-E}
combines pure and forcing evaluations in order to evaluate a bind
fully. An additional cost $c_{\mathsf{let}}$ is added. The rule
\rname{F-R} forces the evaluation of a read expression in the heap
$H$ by first extracting the heap location $l$ from which to read, the
index of the element $n$ to read, and then returning the value stored
in $l$ at index $n$. The rule \rname{F-U} forces the evaluation of an
update expression in a similar way; it returns a unit value. Finally,
the rule \rname{F-L} forces the evaluation of an $\mathsf{alloc}$
expression by creating a new array with the length specified by the
first argument and initial values specified by the second argument,
and by allocating it in the heap at a new location $l$, which is
returned. 
\subsection{Index Terms and Constraints}
In the spirit of DML~\citep{xi99:dependent}, types are indexed
using \emph{static} index terms that are defined in
Figure~\ref{fig:syntax}. Index terms include booleans, natural numbers
and real numbers. A subclass of index terms specific to \THESYSTEM\
and that allows us to reason about arrays is one representing
(potentially infinite) sets of natural numbers. We denote this class
$\beta$. These sets can be used to represent at the type level
different information on arrays. In relational types, they represent
where two related arrays may differ (as explained earlier), while in
unary types, they represent the write permissions for the array. We
will return to this point later, after we explain the types.  We can
explicitly form a set through an indexed set comprehension of the form
$\{I_i \}_{i\in K}$, where $K\subseteq\mathbb{N}$, and we can take the
union $\beta_1\cup\beta_2$ or the difference $\beta_1\setminus\beta_2$
of two sets $\beta_1, \beta_2$.  We consider only
\emph{well-sorted} index terms. To this end, we have a sorting  judgment
of the form $\senv \jiterm{I}{S}$ where $\senv$ is a \emph{sort
environment}, assigning sorts to index variables, and $S$ is a
sort. Our language has five sorts: $\mathbb{N}$ of natural numbers,
used for sizes of arrays; $\mathbb{R}$ of real numbers, used to
express costs; $\mathbb{B}$ of booleans; $\mathbb{P}$ of sets $\beta$ just
described; and, $\mathbb{L}$ of static names $\gamma$ of arrays.  We
omit the sorting judgment because it is straightforward. As a
convention, we use $L$, $U$ to represent unary minimum and maximum
costs, and $D$ to denote a maximum relational cost ($L$, $U$ and $D$
are always of sort $\mathbb{R}$).
%
Index terms can also appear in constraints $C$.
Figure~\ref{fig:syntax} shows some constraints built out of
equalities and inequalities over index terms.


\subsection{Unary and Relational Types}
In {\THESYSTEM} we have two typing modes: \emph{unary}
and \emph{relational}. This separation is also reflected at the type
level where we have two different type languages: \emph{unary types}
$A$ and \emph{relational types} $\tau$.

Unary types ascribe expressions in a single run. They use index terms
to represent size information, as in the case of the type $\lst{I}{A}$
where $I$ represents the size of the list, and costs, as in the case
of the type $\uarrow{A}{{A}'}{(L,U)}$ where $L$ and $U$ represent
lower- and upper-bounds on the cost of the body of the function being
typed. Cost can also be seen as a type system effect. Index terms are
also used for size in basic types like integers, booleans, etc. and
for cost in universal quantification. Additionally, unary types can
contain constraints in types $C\& A$ and $C\supset A$, that can be
used to implement conditional typing.

We also have a type for arrays and a type for impure
computations. The type $\arr{\gamma}{I}{A}$ is the type of arrays of
length $I$ containing objects of type $A$. The annotation $\gamma$
associates a \emph{static name} to the array that is typed. This
static name can be used to refer to the array in other types. Impure
expressions are typed with monadic types. In our case, a monadic unary
type is a cost-annotated Hoare triple type of the shape
$\monadu{P}{\exists \vec{\gamma}.A}{Q}{(L,U)}$, which is inspired by
Hoare Type Theory~\citep{htt}. Assertions $P,Q$ are sets
$\{\gamma_1\to \beta_1,\ldots, \gamma_n\to \beta_n\}$ assigning to
each static location $\gamma_i$ a set of natural numbers $\beta_i$
called the (write) permissions. The idea is that the array named
$\gamma_i$ can be written only at indices in $\beta_i$ (although it
may read anywhere). The index terms $L$ and $U$ are lower- and
upper-bounds on the execution cost of the (forcing) evaluation of the
typed expression.

Relational types ascribe pairs of expressions, one from each of the
two runs and, as we will see in Section~\ref{sec:lr}, they are
actually interpreted as sets of pairs of expressions in our model.  In
relational types, index terms carry not just size information but also
information about the \emph{relation} between the two values from the
two runs.  The type $\lstR{\alpha}{I}{\tau}$ ascribes a pair of lists,
each of length $I$, whose elements are pointwise related at type
$\tau$. Importantly, the relational refinement $\alpha$ specifies an
upper bound on the number of positions at which the corresponding
elements may differ. In other words, at at least $I-\alpha$ positions,
the two lists must have equal elements, even if $\tau$ allows them to
be completely unrelated. The type $\tint[I]$ represents pairs of
integers both of which are equal to $I$. In arrow types
$\rarrow{\tau}{{\tau'}}{(D)}$, the index term $D$ represents an upper
bound on the relative cost of the underlying pair of functions.

Given a pair of unary types $A_1, A_2$, the relational type
$U(A_1,A_2)$ represents arbitrary pairs of expressions of types $A_1$,
$A_2$, respectively. This offers a principle of relationally typing
two ``unrelated'' values. As explained in Section~\ref{sec:overview},
we also have a comonadic relational type $\square \tau$ which
represents pairs of expressions of type $\tau$ which are syntactically
equal. In particular, $\square U(A_1,A_2)$ is the diagonal relation on
$A_1 \cap A_2$.

The relational type $\arr{\gamma}{I}{\tau}$ is similar to the unary
array type but it represents two arrays, each of length $I$,
containing values related at $\tau$ pointwise. $\gamma$ is the static
name for both arrays.  As we will see in Section~\ref{sec:lr}, our
logical relation relates $\gamma$ to two arrays in two different
heaps.  Relational impure computations, illustrated in the
$\mathsf{map}$ example of Section~\ref{sec:overview}, are typed using
a relational cost-annotated monadic type of the form
$\monadR{P}{\exists \vec{\gamma}.\tau}{Q}{D}$. This looks similar to
the unary type
$\monadu{P}{\exists \vec{\gamma}.A}{Q}{(L,U)}$. However, the type
means something very different. In the relational type, the pre- and
post-conditions $P$, $Q$ of form
$\{\gamma_1\to \beta_1,\ldots, \gamma_n\to \beta_n\}$ have a
relational interpretation, namely, that (for all $i$) the two arrays
named $\gamma_i$ must carry equal values at all positions not in
$\beta$ (and the values must be related at $\tau$). At positions in
$\beta$, the values must still be related at $\tau$, but they need not
be equal (unless $\tau$ forces this).  $D$ is an upper bound on the
relative cost of forcing the two impure expressions.

%
As usual, we consider only types that are well-formed.  We have
well-formedness judgments $\lenv;\senv;\cenv \wfa{A}$ for unary types,
and $\lenv;\senv;\cenv \wf{\tau}$ for relational types. Here, $\lenv$
is a \emph{location environment} listing the locations that can appear
in the rest of the judgment, $\senv$ is a sort environment listing all
free index variables, and $\cenv$ is a
\emph{constraint environment} to support conditional typing.

\begin{figure*}
\begin{mathpar}
  
  \inferrule*[right=u-I]
  {  \empty}
  { \lenv ;\senv ; \cenv; \uenv \jtype{0}{0}{n}{\tint[n]} }
\and
 \inferrule*[right=u-f]
  { 
    \inferrule*[]
    {}
    {\lenv ;\senv ; \cenv; x : A, f : \uarrow{A}{B}{(L,U)}, \uenv \jtype{L}{U}{e}{ B }   }
  }
  {\lenv ;\senv ; \cenv; \uenv \jtype{0}{0}{\tfix\, f(x).e }{
  \uarrow{A}{B}{(L,U)} }   }
\and 
  \inferrule*[right=u-V]
  {  \uenv(x)=A}
  {\lenv ; \senv ; \cenv; \uenv \jtype{0}{0}{x}{A} }
\and
  \inferrule*[right=u-t]
  {
    \lenv ; \senv ; \cenv; \uenv  \jtype{L}{U}{t}{ A }  
   \\
   \lenv ;\senv  \wf{P}  
   }
          { \lenv ;\senv ; \cenv; \uenv  \jtype{0}{0}{ \ret{t} }{ \monadu{P}{\exists{\gamma}.A}{P}{(L,U)} }  }
\and
  \inferrule*[right=u-e]
  {
    P=P_1\star P_2 
    \\
    \lenv ; \senv ; \cenv; \uenv  \jtype{L_{1}}{U_{1}}{t_1}{
      \monadu{P_1}{\exists\vec{\gamma_1} .A}{Q_1\star Q_2}{(L,U)} }  
    \\
    \lenv ;  \senv, \vec{\gamma_1} ; \cenv; \uenv , x:A
    \jtype{L_{2}}{U_{2}}{t_2}{ \monadu{Q_1\star P_2}{\exists\vec{\gamma_2} .B}{Q}{(L', U')} }  
    } 
    { \lenv ;\senv ; \cenv; \uenv
      \jtype{L_{1}+L_{2}+L_{l}}{U_{1}+U_{2}+U_{l}}{ \letx{x}{t_1}{t_2} }{
        \monadu{P}{\exists \vec{\gamma_1},\vec{\gamma_2}  .B}{Q\star Q_2}{(L+L', U+U' )} } }
\and
  \inferrule*[right=u-a]
  { 
    \inferrule*[]
    {}
    {\lenv ;\senv ; \cenv; \uenv  \jtype{L_1}{U_1}{t_1}{
        \uarrow{A}{B}{(L,U)} } } \\
    {\lenv ;\senv ; \cenv; \uenv \jtype{L_2}{U_2}{t_2}{A} } 
  }
  {\lenv ;\senv ; \cenv; \uenv \jtype{L_1+L_2+L+L_{app}}{U_1+U_2+U+U_{app}}{\app{t_1}{t_2}}{B} }
\and
  \inferrule*[right=u-l]
  {\lenv ;\senv ; \cenv; \uenv  \jtype{L_1}{U_1}{t_1}{\tint[I]} \\
        \lenv ;  \senv ; \cenv; \uenv  \jtype{L_2}{U_2}{t_2}{A}  \\
        \gamma\ \text{fresh}
        \\
   \lenv ;\senv  \wf{P}  
 }
  {\lenv ;\senv ; \cenv; \uenv  \jtype{0}{0}{\alloc{t_1}{t_2} }{
            \mathrel{\monadu{P}{\exists \gamma . \arr{\gamma}{I}{A}}{
              P\star \gamma \rightarrow \mathbb{N} }{(L_1+L_2 + L_{a}, U_1+U_2+ U_{a})} }  } }
\and
  \inferrule*[right=u-r]
  { 
    {\lenv ;\senv ; \cenv; \uenv \jtype{L_1}{U_1}{ t_1 }{
  \arr{\gamma}{I}{A} }  } \\
  {\lenv ;\senv ; \cenv; \uenv \jtype{L_2}{U_2}{t_2}{\tint[I']}}
  \\  {\senv;\cenv \vDash I' \leq I }
  \\    \lenv ;\senv  \wf{P}  
  }
  {\lenv ;\senv ; \cenv; \uenv \jtype{0}{0}{ \readx{t_1}{t_2 }}{
            \monadu{P}{ \exists \_ . A}{P}{(L_1+L_2+L_r,
              U_1+U_2+U_r) }}    }
\and
  \inferrule*[right=u-u]
  {
    \lenv ;\senv ; \cenv; \uenv \jtype{L_1}{U_1}{t_1}{\arr{\gamma}{I}{A}} 
    \\
    \lenv ;\senv ; \cenv; \uenv \jtype{L_2}{U_2}{t_2}{\tint[I'] }
    \\
    \lenv ;\senv ; \cenv; \uenv \jtype{L_3}{U_3}{t_3}{A}
    \\
    \senv;\cenv \vDash I' \leq I  
    \\
   \lenv ;\senv  \wf{P}  
    \\
        \senv;\cenv \vDash  I' \in \beta  
  }
  {\lenv ;\senv ; \cenv; \uenv \jtype{0}{0}{\updt{t_1}{t_2}{t_3}}{ } 
          {\monadu{ P\star \gamma \rightarrow \beta }{ \exists \_ . \tunit }{P\star \gamma \rightarrow \beta }
        {(L_1+L_2+L_3+L_{u}, U_1+U_2+U_3+U_{u})} } }
\and
\inferrule*[right= u-x ]
  { 
     \lenv ;\senv ; \cenv; \uenv \jtype{L}{U}{t}{A} 
     \\
     \lenv ;\senv; \cenv \subtype{A}{A'} 
     \\
     \senv;\cenv \vDash U \leq U' 
     \\
     \senv;\cenv \vDash L' \leq L  
   }
   {\lenv ;\senv ; \cenv; \uenv \jtype{L'}{U'}{t}{A'}  }

\end{mathpar}
\vspace{0.1in}
\caption{Selection of  unary typing rules.}
\label{fig:unary-typing}
\end{figure*}

\subsection{Unary and Relational Typing} 
\paragraph{Unary Typing Judgments}
\THESYSTEM's unary typing uses the judgment form
$$
\lenv;\senv ; \cenv; \uenv \jtype{L}{U}{t}{A} 
$$
where $t$ is an expression, $\lenv$ is a location environment, $\senv$
is a sort environment, $\cenv$ is a constraint environment, $\uenv$ is
a \emph{unary type environment} assigning unary types to variables,
$A$ is a unary type, and $L$ and $U$ are index terms representing a
lower bound and an upper bound on the cost of evaluating $t$,
respectively. We give a selection of the typing rules for deriving
unary typing judgments in Figure~\ref{fig:unary-typing}. Rules
\rname{U-I} and \rname{U-V} are similar to the
ones available in indexed type systems, with explicit cost $0$.
Rules
\rname{U-F} and \rname{U-A} are similar to the ones available in classical effect
systems. We present them here to show how the costs change in the
typing. 

The remaining rules concern impure expressions of monadic
types. Rules \rname{U-T} and \rname{U-E} type the unit and the bind of
the monad, respectively. They combine the different costs and
assertions in the monadic type, using a style similar to separation
logic. The rule for allocations, \rname{U-L}, introduces a new static
location $\gamma$ and creates a new monadic type whose postcondition
assigns to $\gamma$ all the natural numbers ($\mathbb{N}$), indicating
that all the continuation has the permission to write all positions of
the array. Additionally, like all other rules, this rule also adds a
cost accounting for the forcing of the allocation. Finally, note that
the upper- and lower-bounds on the judgment are $0$. This is because
$\alloc{t_1}{t_2}$ is a value. In the pure evaluation, it returns
without cost. Cost arises only when the term is forced; this is
accounted in the cost annotations in the monadic type. The rule for
reading, \rname{U-R}, merely checks that the index being read is
within the array bounds. The rule for updating,
\rname{U-U}, also performs a similar check but, in addition, it also
requires that the updated index is contained in the permissions
available for the array in the precondition. Finally, the
rule \rname{U-X} allows weakening the upper and lower bounds on the
cost and applying subtyping.

\begin{figure*}
\begin{mathpar}
%
%
\inferrule*[right=r-nc]
  { 
    \lenv ;\senv ; \cenv ; \renv \jtypediff{D}{t}{t}{ \tau}
    \\
    \forall x \in dom(\Gamma).\lenv ; \senv ; \cenv
      \subtype{\renv(x)}{ \square \renv(x) } 
  }
  {\lenv ;\senv ; \cenv; \renv \jtypediff{0}{t}{t}{\square \tau} } 
\and
  \inferrule*[right=r-p]
  { 
    {\lenv ;\senv ; \cenv , C; \renv \jtypediff{D}{t_{1}}{t_{2}}{ \tau}}
    \\
    {\lenv ; \senv ; \cenv, \neg C; \renv \jtypediff{D}{t_1}{t_2}{\tau}}
  }
  {\lenv ;\senv ; \cenv; \renv \jtypediff{D}{t_1}{t_2}{\tau} } 
\end{mathpar}
\vspace{0.1in}
\caption{Selection of pure  relational synchronous typing rules.}
\label{fig:relational-syn-typing-pure}
\end{figure*}

\begin{figure*}
\begin{mathpar}
 \inferrule*[right=r-s]
  { 
    \inferrule*[]
    {}
    {\lenv ;\senv ; \cenv ; |\renv|_1 \jtype{L_1}{U_1}{t_1}{A_1}}
    \\
    \inferrule*[]
    {}
    {\lenv ;\senv ; \cenv; |\renv|_2 \jtype{L_2}{U_2}{t_2}{A_2}}
  }
  {\lenv ;\senv ; \cenv; \renv
    \jtypediff{U_1-L_2}{t_1}{t_2}{U(A_1,A_2)} } 
\and
\inferrule*[right=r-lt-e]
  { 
    \inferrule*[]
    {}
    {\lenv ;\senv ; \cenv ; |\renv|_1 \jtype{L_1}{U_1}{t_1}{  A_1    }}
    \and
    \inferrule*[]
    {}
    {\lenv ;\senv ; \cenv; \renv,x:U(A_1,A_1) \jtypediff{D_2}{t_2}{t_2'}{\tau}  }
  }
  {\lenv ;\senv ; \cenv; \renv  \jtypediff{U_1+D_2 +c_{lt} }{\letm{x}{t_1}{t_2} }{t_2'}{\tau} } 
\and
\inferrule*[right=r-e-lt]
  { 
    \inferrule*[]
    {}
    {\lenv ;\senv ; \cenv ; |\renv|_2 \jtype{L_1}{U_1}{t_1'}{ A_1'   }}
    \and
    \inferrule*[]
    {}
    {\lenv ;\senv ; \cenv; \renv,x: U(A_1',A_1') \jtypediff{D_2}{t_2}{t_2'}{\tau'}  }
  }
  {\lenv ;\senv ; \cenv; \renv  \jtypediff{D_2 - L_1 - c_{lt} }{t_2   }{\letm{x}{t_1'}{t_2'} }{\tau'} } 
\and
\inferrule*[right=r-app-e]
  { 
    \inferrule*[]
    {}
    {\lenv ;\senv ; \cenv ; |\renv|_1 \jtype{L_1}{U_1}{t_1}{
        \uarrow{A_1}{{A_2}}{(L,U)}    } }
    \and
    \inferrule*[]
    {}
    {\lenv ;\senv ; \cenv; \renv \jtypediff{D_2}{t_2}{t_2'}{U(A_1, A_2')}  }
  }
  {\lenv ;\senv ; \cenv; \renv  \jtypediff{U_1+U+D_2 +c_{app} }{t_1  \eapp t_2 }{t_2'}{U(A_2,A_2')} } 
%
\and
\inferrule*[right=r-case-e]
  {  
    \inferrule*[]
    {}
     {\lenv ;\senv ; \cenv ; |\renv|_1 \jtype{L_1}{U_1}{t}{
           A_1 + A_2 } }
    \\
    \inferrule*[]
    {}
    {\lenv ; \senv ; \cenv; \renv ,x:U(A_1,A_1) \jtypediff{D_2}{t_1}{t'}{\tau}}
    \\
    \inferrule*[]
    {}
    {\lenv ; \senv ; \cenv; \renv ,y:U(A_2,A_2) \jtypediff{D_2}{t_2}{t'}{\tau}}
   }
  {\lenv ; \senv ; \cenv; \renv \jtypediff{U_1+D_2+c_{case}}{
      \ecase(t,x.t_1,y.t_2) }{t'}{\tau}}
%
%
\end{mathpar}
\vspace{0.1in}
\caption{Selection of pure relational asynchronous typing rules.}
\label{fig:relational-asyn-typing-pure}
\end{figure*}

\paragraph{Relational Typing Judgments} 
\THESYSTEM's relational typing uses the judgment form
$$
\lenv; \senv ; \cenv; \renv \jtypediff{D}{t_1}{t_2}{\tau}
$$
Here, $t_1$ and $t_2$ are two expressions, $\lenv$, $\senv$, and
$\cenv$ are environments similar to the ones used by unary typing
judgments, $\renv$ is a \emph{relational type environment} assigning
relational types to variables, $\tau$ is a relational type for $t_1,
t_2$, and $D$ is an index term representing an \emph{upper bound} on
the relative cost of evaluating $t_1$ and $t_2$, i.e.,
$\text{cost}(t_1)-\text{cost}(t_2)$. \rd{We have two kinds of relational typing rules:
\emph{synchronous rules} which relate two structually similar programs, and
\emph{asynchronous rules} which relate programs that are not
necessarily structurally similar. We 
first present a selection of the \emph{pure} typing rules which include both synchronous
rules (Figure~\ref{fig:relational-syn-typing-pure})
and asynchronous rules (Figure~\ref{fig:relational-asyn-typing-pure}), inspired by the work of
\citet{Cicek17}. Then, we present a selection of the \emph{monadic}
typing rules which support relational cost analysis for arrays. We
present the synchronous rules
(Figure~\ref{fig:relational-typing-syn-impure}) and the
asynchronous rules
(Figure~\ref{fig:relational-typing-asyn-impure}). The rest of the
typing rules can be found in the Appendix.}

\rd{\paragraph{Pure Synchronous Rules} We present two synchronous
rules \rname{R-P} and \rname{R-NC} in
Figure~\ref{fig:relational-syn-typing-pure}. The other rules are
similar to the one from~\citet{Cicek17} and they can be found in the Appendix.} Rule \rname{R-P} allows
reasoning by cases on any constraint in the constraint environment.
Rule \rname{R-NC} is the introduction rule for $\square$-ed
types. Briefly, $t$ can be related to itself at type $\square \tau$
when $t$ relates to itself at type $\tau$ and, additionally, all
variables in the context morally have $\square$-ed types. The latter
ensures that variables can only be substituted by equal terms. In this
case, the relative cost is trivially $0$. 
  \rd{\paragraph{Pure Asynchronous Rules} We present a selection of
    the pure asynchronous rules in
    Figure~\ref{fig:relational-asyn-typing-pure} including the generic
    rule \rname{R-S}, and rules for the pure let binding, function application and case elimination.}
 Rule \rname{R-S} allows
switching from relational reasoning about $t_1$ and $t_2$ to unary
reasoning about the two terms, independently. Notice that the
relational type in the conclusion is the embedding of the two unary
types without any meaningful relation ($U(A_1,A_2)$). The rule uses a
straightforward map $|\renv |_i$ from relational environments to unary
environments, whose definition can be found in the
appendix. Importantly, the relative cost in the conclusion is the
difference of the unary costs in the premises.
 \rd{ Rule \rname{R-LT-E} relates a pure let binding
expression to an arbitrary expression. In this rule, we use
the metavariable $c_{lt}$ to denote the cost of a let
elimination. Notice that one of the assumptions in this rule, the one
for the expression $t_1$, is a unary typing judgment. This is needed to
provide guarantees on typability of $t_1$ and to provide the cost of
evaluating it, which is used in the bound on the relative cost in the
conclusion of the rule.   The rule \rname{R-E-LT} is dual to
\rname{R-LT-E} -- it relates an arbitrary expression with a let. Notice
that while  the rule \rname{R-LT-E} uses the upper bound on the unary
cost of $t_1$, the rule \rname{R-E-LT} uses the lower bound. 
Rule \rname{R-APP-E} relates a function application with an arbitrary
expression, while rule \rname{R-CASE-E} relates a case expression with
an arbitrary expression. Also in these rules we use some unary typing
assumptions to guarantee typability and to provide unary costs which
are used in giving upper bounds on the relative costs. We also have dual
rules which we present in the Appendix. }

\begin{figure*}
\begin{mathpar}
  \inferrule*[right=r-let]
  {
          P=P_1\star P_2
          \\
    {\lenv ;\senv ; \cenv; \renv \cmp{t_{1}}{t_{1}'}{
                    \monadR{P_1}{\exists\vec{\gamma_1}.\tau}{Q_1\star Q_2}{D}}{D_1}  }
    \\
    {\lenv ;\senv,\vec{\gamma_1}:\vec{\mathbb{L}};\cenv;\renv,x:\tau\cmp{t_2}{t_2'}{\monadR{Q_1\star
                      P_2}{\exists\vec{\gamma_2}.\sigma}{Q}{D'}}{D_2}  }
    } 
  {\lenv ;\senv ; \cenv; \renv  \cmp{ \letx{x}{t_{1}}{t_{2}} }{
            \letx{x}{t_{1}'}{t_{2}'}}{
            \monadR{P}{\exists\vec{\gamma_1}\vec{\gamma_2}
              .\sigma}{Q\star Q_2}{D+D'}
          }{D_1+D_2} }
\and
\inferrule*[right=r-t]
  {
    {\lenv ;\senv ; \cenv; \renv  \cmp{t_1}{t_2}{\tau}{D}  }
    \\ 
   \lenv ;\senv  \wf{P}  
    } 
  {\lenv ;\senv ; \cenv; \renv \cmp{\ret{t_1}}{\ret{t_2}}{ { \monadR{P}{\exists\_.\tau}{P}{D} }  }{0}}
\and
  \inferrule*[right=r-l]
  {
    {\lenv ;\senv ; \cenv; \renv  \jtypediff{D_1}{t_1}{t_1'}{\tint[I]} }
   \quad
    {\lenv ;\senv ; \cenv; \renv  \jtypediff{D_2}{t_{2}}{t_2'}{\tau}} 
    \quad
\gamma\ \mathsf{fresh}
     \\    
\lenv ;\senv  \wf{P}  
   }
  {\lenv ;\senv ; \cenv; \renv \jtypediff{0}{\alloc{t_{1}}{t_2} }{\alloc{t_1'}{t_2'}}{ \mathrel{  \monadR{P}{\exists \gamma.\arrR{\gamma}{I}{\tau}}{P\star \gamma\rightarrow \mathbb{N} }{D_1+D_2}  } } }
\and
\inferrule*[right=r-lb]
  {    {\lenv ;\senv ; \cenv; \renv  \jtypediff{D_1}{t_1}{t_1'}{\tint[I]} }
     \\
   {\lenv ;\senv ; \cenv; \renv
   \jtypediff{D_2}{t_{2}}{t_2'}{\tbox\tau}} 
   \\
  \gamma\ \mathsf{fresh}
   \\
   \lenv ;\senv  \wf{P}  
    }
   {\lenv ;\senv ; \cenv; \renv \jtypediff{0}{\alloc{t_{1}}{t_2}
     }{\alloc{t_1'}{t_2'}}{ \mathrel{  \monadR{P}{\exists
           \gamma.\arrR{\gamma}{I}{\tau}}{ P\star \gamma \rightarrow \emptyset  }{D_1+D_2}  } } }
\and
%
%
%
%
\inferrule*[right=r-r]
  { 
   {\lenv ; \senv ; \cenv; \renv \jtypediff{D_1}{t_{1}}{t_{1}'}{ {\arrR{\gamma}{I}{\tau} }  } }
    \\
    { \lenv ;\senv ; \cenv; \renv
      \jtypediff{D_2}{t_{2}}{t_2'}{\tint[I']} }
   \\
    { \senv; \cenv \models I'\leq I }
   \\
    \lenv ;\senv  \wf{P}  
  }
  {\lenv ;\senv ; \cenv; \renv \jtypediff{0}{\readx{t_{1}}{t_{2}} }{ \readx{t_{1}'}{t_{2}'} }{    
            \monadR{P}{ \exists \_ .\tau  }{P}{D_1+D_2} }} 
%
%
%
\and
  \inferrule*[right=r-rb]
  { 
   {\lenv ;\senv ;\cenv;\renv\jtypediff{D_1}{t_{1}}{t_{1}'}{\arrR{\gamma}{I}{\tau}}}
    \\
    { \lenv ;\senv ; \cenv; \renv \jtypediff{D_2}{t_{2}}{t_2'}{\tint[I']} }
    \\
    {\senv; \cenv \models I'\leq I }
                \\
    { I' \not\in \beta }
    \\
\lenv ;\senv  \wf{P}  
  }
  {\lenv ;\senv ; \cenv; \renv \jtypediff{0}{\readx{t_{1}}{t_{2}} }{ \readx{t_{1}'}{t_{2}'} }{    
            \monadR{P\star \gamma\mapsto\beta}{ \exists \_ .\tbox \tau  }{P\star\gamma\mapsto \beta}{D_1+D_2} }} 
\and
  \inferrule*[right=r-u]
  {
    {\lenv ;\senv ; \cenv; \renv \jtypediff{D_1}{t_1}{t_1'}{\arrR{\gamma}{I}{\tau}}} 
     \\
    {\lenv ;\senv ; \cenv; \renv \jtypediff{D_2}{t_{2}}{t_{2}'}{\tint[I']}} 
   \\
    {\lenv ;\senv ; \cenv; \renv \jtypediff{D_3}{t_{3}}{t_{3}'}{\tau}} 
   \\
    {\senv ; \cenv \models I'\leq I}
    \\
    \lenv ;\senv  \wf{P}  
  }
  {\lenv ;\senv ; \cenv; \renv \jtypediff{0}{\updt{t_{1}}{t_{2}}{t_{3}} }{  \updt{t_{1}'}{t_{2}'}{t_{3}'} }{ 
\monadR{P\star\gamma\mapsto\beta}{ \exists\_ . \trunit}{P\star\gamma\mapsto \beta \cup \{ I' \} }{D_1+D_2+D_3}}} 
\and
\inferrule*[right=r-ub]
  {
    {\lenv ;\senv ; \cenv; \renv \jtypediff{D_1}{t_{1}}{t_{1}'}{\arrR{\gamma}{I}{\tau}}} 
     \\
    {\lenv ;\senv ; \cenv; \renv \jtypediff{D_2}{t_{2}}{t_{2}'}{\tint[I']}} 
    \\
    {\lenv ;\senv ; \cenv; \renv \jtypediff{D_3}{t_{3}}{t_{3}'}{\tbox\tau}} 
    \\
    {\senv ; \cenv \models I'\leq I}
    \\
    \lenv ;\senv  \wf{P}  
  }
  {\lenv ;\senv ; \cenv; \renv \jtypediff{0}{\updt{t_{1}}{t_{2}}{t_{3}} }{  \updt{t_{1}'}{t_{2}'}{t_{3}'} }{ 
\monadR{P\star \gamma\mapsto\beta}{ \exists\_ .
  \trunit}{P\star\gamma\mapsto \beta  \setminus \{ I' \} }{D_1+D_2+D_3}}} 
\and
\inferrule*[right=r-fix-ext]
  {
  {\lenv ;\senv ; \cenv;  x: \tau_1 , f \ :  \rarrow{\tau_1}{\tau_2}{(D)},\renv , f \ : U (A_1,A_2)   \jtypediff{D}{t_1}{t_2}{\tau_2 } }
  \\
  {\lenv ;\senv ; \cenv;  |\renv|_1 \jtype{0}{0}{\tfix \, f(x). t_1}{A_1} }
  \\
  {\lenv ;\senv ; \cenv;   |\renv|_2 \jtype{0}{0}{\tfix \, f(x). t_2}{A_2} }
  }
  { \lenv ;\senv ; \cenv; \renv \jtypediff{D}{ \tfix \, f(x). t_1 }{ \tfix \, f(x). t_2 }{\rarrow{\tau_1}{\tau_2 }{(D)} }}
\end{mathpar}
\caption{Selection of monadic synchronous relational typing rules.}
\label{fig:relational-typing-syn-impure}
\end{figure*}

\begin{figure*}
\begin{mathpar}
\inferrule*[right=r-let-e]
  { 
    \inferrule*[]
    {}
    {\lenv ;\senv ; \cenv ; |\renv|_1 \jtype{L_1}{U_1}{t_1}{  
   \monadu{P_1}{\exists\vec{\gamma_1} :A_1}{Q_1}{(L,U)}
    } }
\and
     \inferrule*[]
    {}
    {\lenv ;\senv ; \cenv ; |\renv|_2 \jtype{L_2}{U_2}{t_2'}{ 
   \monadu{P_2}{\exists\vec{\gamma_1} :A_1'}{Q_2}{(L',U')}
   }}
    \\\\
    \inferrule*[]
    {}
    {dom(P)=dom(P_1)}
     \and    
\inferrule*[]
    {}
    {\lenv ;\senv ; \cenv; \renv,x:U(A_1,A_1)
      \jtypediff{D_2}{t_2}{t_2'}{
    \monadR{P \sqcup P_1}{\exists\vec{\gamma_1}.\tau}{Q }{D'}
}  }
  }
  {\lenv ;\senv ; \cenv; \renv  \jtypediff{-L_2
    }{\letx{x}{t_1}{t_2} }{t_2'}{
   \monadR{P}{\exists\vec{\gamma_1}.\tau}{Q }{U_1+U+(D_2+U_2) +D' +c_{let}}
} } 
\and
\inferrule*[right=r-e-let]
  { 
    \inferrule*[]
    {}
    {\lenv ;\senv ; \cenv ; |\renv|_2 \jtype{L_1}{U_1}{t_1'}{ 
   \monadu{P_1}{\exists\vec{\gamma_1} :A_1'}{Q_1}{(L,U)}
   }}
    \and
     \inferrule*[]
    {}
    {\lenv ;\senv ; \cenv ; |\renv|_1 \jtype{L_2}{U_2}{t_2}{ 
   \monadu{P_2}{\exists\vec{\gamma_1} :A_1}{Q_2}{(L',U')}
   }}
    \\\\ 
   \inferrule*[]
    {}
    {dom(P)=dom(P_1)}
     \and
    \inferrule*[]
    {}
    {\lenv ;\senv ; \cenv; \renv,x: U(A_1',A_1')
      \jtypediff{D_2}{t_2}{t_2'}{
\monadR{P \sqcup P_1}{\exists\vec{\gamma_1}.\tau'}{Q }{D'}
}  }
  }
  {\lenv ;\senv ; \cenv; \renv  \jtypediff{ U_2
    }{t_2   }{\letx{x}{t_1'}{t_2'} }{
\monadR{P}{\exists\vec{\gamma_1}.\tau'}{Q }{D' +(D_2-L_2) - L_1 -L -
  c_{\mathrm{let}} }
} } 
\end{mathpar}
\caption{Selection of monadic asynchronous relational typing rules.}
\label{fig:relational-typing-asyn-impure}
\end{figure*}

\paragraph{Synchronous Rules}
Figure~\ref{fig:relational-typing-syn-impure} shows a selection of
relational synchronous
typing rules pertaining to monadic constructs and arrays.
Rules \rname{R-T} and \rname{R-LET} relationally type the return and
bind of our monad. The rules introduce the trivial relational
Hoare-triple and combine two relational Hoare triples by sequencing,
respectively. In particular, the rule \rname{R-LET} uses the style of
separation logic.  Rule \rname{R-FIX-EXT} types fixpoint expressions
relationally. During the relational reasoning, it also allows assuming
the \emph{unary} types of the two functions, which are established in
separate premises. This rule, introduces a weak form
of \emph{intersection types} in the environment which can be used in
combination with the rule \rname{R-S}
(Figure~\ref{fig:relational-asyn-typing-pure}) to give precise bounds on
relative cost.

For each operation on arrays we have two rules, one that is general
and the other that works under some assumption about equality of
arguments in the two runs. Consider, for example, the
rules \rname{R-L} and \rname{R-LB} for relationally typing the
$\mathsf{alloc}$ construct. The rules are similar, e.g., both create a
new static name $\gamma$ for the two allocated arrays and both account
for relative costs very similarly. However, \rname{R-LB} applies only
when the expressions initializing the two arrays are related at a
$\square$-ed type (second premise). As a result, it is guaranteed that
the arrays allocated in the two runs will have equal values in all
positions. This is reflected in the assertion
$\gamma \rightarrow \emptyset$ in the postcondition of the monadic
type in the rule, which says that there are no locations where the
newly allocated arrays (named $\gamma$) can differ. In contrast, the
rule \rname{R-L} does not require the initializing expressions to be
related at a $\square$-ed type, but it has
$\gamma \rightarrow \mathbb{N}$ in the postcondition, meaning that the
two arrays may differ anywhere. A similar difference arises in the
rules \rname{R-R} and \rname{R-RB} for relationally typing the
construct $\mathsf{read}$. In \rname{R-RB}, the read index $I'$ must
not be in the $\beta$ of the array being read in the precondition; as
a result, the values read must be equal in the two runs. Hence, the
resulting type has a $\square$ on it. \rname{R-R} is similar, but,
here, there is no requirement that $I'$ is not in the $\beta$, so two
different values may be read, and there is no $\square$ on the result
type. The rules \rname{R-U} and \rname{R-UB} for $\mathsf{updt}$
follow the principle of $\mathsf{alloc}$: In \rname{R-UB}, the values
being written in the two runs are known to be equal (via a
$\square$-ed type), so the index $I'$ that is updated is removed from
$\beta$ in the postcondition. This is not the case in \rname{R-R},
where it must be added to $\beta$, since the two values at index $I'$
might differ after the update.\footnote{The astute reader will note
that the set of $\gamma$s in any pre- or postcondition must be written
down explicitly, i.e., we have not introduced sophisticated
constructors (like set comprehension) for pre- and
postconditions. This means that we cannot meaningfully specify monadic
computations that allocate a data-dependent number of arrays. This
hasn't been a problem for our examples, and we believe an extension to
lift this restriction will be straightforward.} \rd{In all these rules, the premise $
\senv; \cenv \vDash I' \leq I$ denotes a constraint entailment which
reads as follows: under the
substitution of all the variables in the index environment $\senv$,
under the assumption of the constraint $\cenv$, the constraint $I'
\leq I$ holds. This premise gurantees that the array bound is not
exceeded. We omit here the rules for deriving this judgment since they
are standard.} 

Finally, note that all monadic rules ``propagate'' relative costs from
the premises to the monadic types. This is similar to the unary rules;
the difference is that the costs propagated here are relative, whereas
the unary type system propagates unary lower- and upper-bounds.

\paragraph{Asynchronous Rules}
\rd{ Figure~\ref{fig:relational-typing-asyn-impure} shows the two
  asynchronous rules \rname{R-LET-E} and \rname{R-E-LET}, relating a
  monadic binding construct and an arbitrary expression. We explain
  only the rule \rname{R-LET-E} which relates the monadic binding
  construct $\letx{x}{t_1}{t_2}$ to an arbitrary expression $t_2'$ 
  (the rule \rname{R-E-LET} is its dual and it can be
  understood similarly). 
  The first premise of the rule \rname{R-LET-E} requires a unary typing for the
  monadic expression $t_1$. This typing has two kinds of costs: the
  lower bound $L_1$ and upper bound $U_1$ for the unary execution cost
  of $t_1$, and the lower bound $L$ and upper bound $U$ for the
  execution cost of the resulting computation evaluated from $t_1$,
  this is embedded in the monadic type of $t_1$. 
%
  The second premise requires a unary typing for the
  monadic expression $t_2'$. This gives us an upper bound $U_2$ on the
  cost of evaluating this expression.
  The premise $dom(P) =dom( P_1)$ requires that
  the execution of the computation resulting from the expression $t_1$
  can only affect arrays that appear
  in both $P_1$ and $P$.
  Finally, the last premise requires relating the subexpression $t_2$ to
  $t_2'$ with the relative cost upper-bounded by $D_2$ under the
  assumption that the values substituted for the variable $x$ are related at the type
  $U(A_1,A_1)$. Notice that this is the weakest requirement in terms
  of types that we can have. Additionally, this typing judgment also
  gives us the upper bound $D'$ on the relative cost for executing the
  two computations resulting from evaluating the two expressions. 
To put the information of the unary and
  relational typing together we use the
  precondition $P\sqcup P_1$ in this premise, where the operation
  $\sqcup$  gives a precondition where a name $\gamma$ which is
  used in $P$, e.g. $\gamma\mapsto\beta\in P$, and in $P_1$,
  e.g. $\gamma\mapsto\beta_1\in P_1$, now points to the union of the two
  corresponding sets, i.e. e.g. $\gamma\mapsto\beta\cup\beta_1\in
  P\sqcup P_1$.
The conclusion of the rule uses all the cost
  information we discussed to compute an upper bound on the relative
  cost of the two expressions, where, as usual, we use the
  metavariable $c_{let}$ to denote the 
  cost of evaluating the monadic binding construct.}

\rd{One can also design similar asynchronous rules for the other monadic
constructs. However, the syntactic forms of the other
constructs considerably constrain their asynchronous typing rules, making the scope of application of such rules rather
narrow. For this reason we do not commit to the design of such rules here.  
}

\paragraph{Subtyping}
Subtyping is important in \THESYSTEM. It serves several
purposes. First, as in all refinement type systems, subtyping equates
terms up to refinement, e.g., it allows replacing $\tint[2+i]$ with
$\tint[5]$ under the constraint $i = 3$. Second, specific to cost
analysis, subtyping allows weakening costs, e.g., the relational type
$\rarrow{\tau_1}{\tau_2}{(D)}$ can be subtyped to
$\rarrow{\tau_1}{\tau_2}{(D')}$ when $D \leq D'$ since the $D$ on the
arrow is an upper bound on relative cost. Third, subtyping allows
``massaging'' of modalities $\square$ and $U$, e.g., $\square \tau$
can be subtyped to $\tau$. Finally, specific to \THESYSTEM, subtyping
allows weakening of pre- and postconditions in monadic types. The
first three uses are standard (e.g., see~\citet{Cicek17}), so we only
describe the last use here. The unary and relational subtyping
judgments have the forms $\lenv;\senv ; \cenv \subtypeA{A_{1} }{ A_{2}
}$ and $\lenv;\senv ; \cenv \subtype{ \tau_{1} }{ \tau_{2} } $,
respectively.  Figure~\ref{fig:subtyping} shows selected subtyping
rules. The notation $P \subseteq P'$ means that
$P=\{ \gamma_1 \rightarrow \beta_1,\gamma_2 \rightarrow \beta_2, \dots, \gamma_n \rightarrow \beta_n \}$,
$P'= \{ \gamma_1 \rightarrow \beta_1',\gamma_2 \rightarrow \beta_2' ,
\dots, \gamma_n \rightarrow \beta_n'  \}$, and  $\forall i \in \{ 1,\dots,n \} . \beta_i \subseteq \beta_i'$.

Rule \rname{S-UM} allows subtyping on the unary monadic type. It says
that we can subtype by weakening the costs, adding more (write)
permissions to the pre-condition and removing permissions from the
postcondition, as manifest in the premises $P \subseteq P'$ and
$Q' \subseteq Q$. Rule \rname{S-RM} similarly allows subtyping on the
relational monadic type. This rule says that we can subtype by
weakening the relative cost, making the precondition more precise and
the postcondition less precise, where $P'$ is more precise than $P$
when $P'$ tells us more about which values are equal. In particular,
$\gamma \rightarrow \beta$ is more precise than
$\gamma \rightarrow \beta'$ when $\beta' \subseteq \beta$. This is why
the premises of \rname{S-RM} check $P' \subseteq P$ and $Q \subseteq
Q'$. Note that the checks on $P, P'$ and $Q, Q'$ are dual in the two
rules. This is pure coincidence; the meanings of the
pre-(post-)condition in the unary and relational monadic types are
completely different. Finally, rule \rname{S-RUM} allows subtyping
from $U$ applied to two unary monadic types to a single relational
monadic type. This rule is best read as follows: If we have two
computations that modify an array ($\gamma_i$) at positions in $T_i$
and $T_i'$, respectively (left side of $\sqsubseteq$), then running
them on two arrays that agree at all positions outside the set $\beta$
will result in two arrays that agree at all positions outside the set
$\beta \cup T_i \cup T_i$' (right side of $\sqsubseteq$).

\begin{figure*}
  \begin{mathpar}
  %
  %
   \inferrule*[right = s-um ]
  { \begin{array}{c}
    {\lenv ; \senv ; \cenv \subtypeA{A}{A'} } \quad  { \vec{\gamma_1} \subseteq \vec{\gamma_2} }
    \arcr
    { \senv ; \cenv \subcost{L'}{L}{\leq} }
     \quad
    { \senv ; \cenv \subcost{ U }{ U' }{ \leq} }
\end{array}
\quad
\begin{array}{c}
    { \lenv,\senv ;\cenv  \subcost{P}{P'}{ \subseteq } } 
\arcr
    { \lenv, \vec{\gamma_1};\senv;\cenv  \subcost{Q'}{Q}{ \subseteq }  }
\end{array}
  }
  {\lenv ; \senv ; \cenv \subtypeA{ \monadu{P}{ \exists \vec{\gamma_1} . A}{Q}{(L,U)} }{ \monadu{P'}{\exists \vec{\gamma_2}. A'}{Q'}{(L',U')} } }
  %
  %
  %
  %
  \and
 \inferrule*[right = s-rm ]
  { 
    { \lenv ;\senv ; \cenv \subtype{\tau}{\tau'} }
    \quad
    { \senv ; \cenv \subcost{D}{D'}{\leq} }
   \\
    {\lenv, \vec{\gamma_1}; \senv; \cenv \subcost{Q}{Q'}{ \subseteq } } 
     \\
       { \lenv; \senv ;\cenv  \subcost{P'}{P}{ \subseteq } } 
   \\ { \vec{\gamma_1} \subseteq \vec{\gamma_2} }
  }
  {\lenv ; \senv ; \cenv \subtype{ \monadR{P}{\exists \vec{\gamma_1}.\tau}{Q}{D} }{ \monadR{P'}{ \exists \vec{\gamma_2} . \tau'}{Q'}{D'} } }
\vspace{-3mm}
\\
\inferrule*[ lab = s-rum]
  { \beta_i ' = \beta_i \cup T_i \cup T'_i  }
  { \lenv ; \senv ; \cenv  \vDash {  U ( \monadu{\gamma_i  \Arrow{.2cm} 
  T_i}{\exists \vec{\gamma_1'}. A_1}{Q_1}{(L,U)} , \monadu{\gamma_i
  \Arrow{.2cm} T'_i}{\exists \vec{\gamma_2'}. A_2}{Q_2}{(L',U')} ) } 
  \sqsubseteq {  \monadR{\gamma_i  \Arrow{.2cm}  \beta_i }{
      \exists{\vec{\gamma_1},\vec{\gamma_2}} . U(A_1,A_2) }{ \gamma_i
       \Arrow{.2cm}  \beta_i'  }{ U - L' }
  } }
\vspace{-2mm}
\end{mathpar}
\caption{Selection of subtyping rules.}
\label{fig:subtyping}
\end{figure*}

\section{Logical Relations}
\label{sec:lr}
To prove the soundness of \THESYSTEM\ we build a step-indexed logical
relation for its types. We give two interpretations---one unary and one monadic, that interact at the type $U(A_1, A_2)$.

\paragraph{Unary Interpretation}
The \emph{value interpretation} $\llu{A}{g,k} $ of a unary type $A$
is, as usual, a set of values. Also, as usual, this interpretation is
indexed by a ``step-index'' $k\in \mathbb{N}$, which is merely a proof
device for induction~\cite{ahmed06:sll,AppelM01:si}. The step index
counts the ``steps'' in our operational semantics. Importantly, the
interpretation is also refined by a mapping $g$ from static names
$\gamma$ to triples $(l,n,A)$ expressing the location, the length of
the array, and the \emph{syntactic} type of the elements of the array
named $\gamma$. $g$ is our version of a Kripke world from the literature
on logical
relations~\cite{DBLP:journals/jfp/NeisDR11,DBLP:conf/popl/TuronTABD13}. We
give a selection of the clauses defining the value interpretation of
unary types in Figure \ref{fig:unary-interpret}.

For example, the value interpretation $\llu{\arr{\gamma}{n}{A}}{g,k}$
of an array type is a set of locations $l$ which are assigned to
$\gamma$ in $g$.  The value interpretation
$\llb{ \monadu{P}{\exists{\vec{\gamma}}.A}{Q}{(L,U)} }_{g,k}$ of a
monadic type uses a heap relation $H\models_{g,k} P$ which says that
the assertion $P$ holds for the heap $H$ at world $g$. All the static
names $\gamma$ in $P$ refer through $g$ to concrete arrays in $H$ that
have the right type and length.
The value interpretation for monadic types is a set of monadic values
$v$ that when forced using a heap $H$ validating the precondition $P$,
yield a heap $H_1$ validating the postcondition $Q$. Additionally, the
interpretation only allows those computations $v$ that update arrays
at locations for which the precondition $P$ asserts permissions.  We
can extend the value interpretation to expressions:
 $$\llu{A}{g,k}^{E,(L,U)} = \{ t  \, | \, \forall v, k'\leq  k. \eval{t}{v}{c,k'} \; \Rightarrow v \in \llu{A}{g,k-k'}  \land L \leq c \leq U \} $$
This definition accounts for costs.
%
The interpretation is also extended to environments in a standard way.
In the following theorem we use $\vdash \delta: \Delta$ and $\vDash \delta \Phi $  to denote that $\delta$ is a substitution for the index variables in $\Delta$ satisfying the constraint environment $\Phi$. 

\begin{theorem}[Fundamental Theorem for Unary Typing]
If $\lenv; \senv; \cenv; \uenv \vdash_{L}^U t: A $, $\vdash \delta: \Delta$ and  $ \vDash \delta \Phi $,  and  
$ \sigma \in \llu{\delta \Omega}{g,k} $,
then $ (\delta \sigma t ) \in \llu{ \delta A}{g,k}^{E,(\delta L, \delta U)} $.
\end{theorem}

\begin{figure*}
\begin{mathpar}
  \llu{ \tint[n]}{g,k} = \{ \,  n\, \}  
\quad
\llp{\trint[n] }{G,k}   = \{ \, ( n,n) \}  
  \quad
  \llu{\arr{\gamma}{n}{A}}{g,k}    = \{ \,  l \, | \, g(\gamma)=(l,n,A)    \}
\\
      \llp{\arrR{\gamma}{n}{\tau}}{G,k}    = \{ \, ( l_1,l_2) \, | \, G(\gamma)=(l_1,l_2,\tau,n) \} 
%
  \quad
 \llp{\square \tau }{G,k}   = \{ \, ( v, v ) \, | \, (v,v) \in \llp{\tau}{G,k}  \} 
\\
      \llp{ U(A_1,A_2)}{G,k}    = \{ \, ( v_1,v_2) \, | \, \forall k', v_1 \in \llu{A_1}{G|_1, k'} \land v_2 \in \llu{A_2}{G|_2,k'}  \}  
\\
      \llb{ \monadu{P}{\exists{\vec{\gamma}}. A}{Q}{(L,U)} }_{g,k}  = \left \{\begin{array}{l} v \, | \,  \forall g_1\supseteq g,\forall k_1\leq k,\forall k_2 < k_1, c,  \forall H. \big  (H \vDash_{g_1,k_1} P \land v;H\Downarrow_f^{c,k_2}\big )\\  \Rightarrow   \exists g_2\supseteq g_1,H_1,v_1,\vec{\gamma}. 
( \evalf{ { v;H } }{ { v_1;H_1} }{c,k_2} )  \land L \leq c \leq U   \land  H_1 \vDash_{g_2,k_1-k_2} Q \\ \land    v_1 \in  \llu{A}{g_2,k_1-k_2}   \land \big ( (\exists n. P = \{ \gamma_1 \rightarrow T_1,\dots, \gamma_n \rightarrow T_n \} \\  \land \forall i \in [1,n]. g(\gamma_i)= (l_i,A,m)) \Rightarrow   \forall j. (H[l_i][j]) \not = H_1[l_i][j] \Rightarrow j \in T_i \big ) \end{array}\right \}   
\vspace{-2mm}
\\
{ \llp{\monadR{P}{\exists \vec{\gamma}.\tau}{Q}{D} }{G,k} =\left
    \{ \begin{array}{l}  (v_1,v_2) \, | \, \forall G_1\supseteq G, k_1
         \leq k, k_2 < k_1,k_3,c_1,c_2, H_1,H_2.\\ \Big ({(H_1,H_2)}
         \vDash_{G_1,k_1} P  \land   v_1;H_1\Downarrow_f^{c_1,k_2}
         \land v_2;H_2\Downarrow_f^{c_2,k_3} \Big ) \Rightarrow \\ \exists G_2 \supseteq G_1, H_1',H_2', v_1',
v_2',\vec{\gamma}.  \Big ( \evalf{ v_1;H_1 }{ { v_1';H_1'} }{c_1,k_2}  \land \, \evalf{ {v_2;H_2} }{ {v_2';H_2'} }{c_2,k_3} \\ \land  {(H_1',H_2') } \vDash_{{G_2,k_1-k_2}} Q  \land \, {(v_1',v_2')} \in  \llp{  \tau }{G_2,k_1-k_2} \land c_1-c_2 \leq D \Big ) 
\end{array} \right \}}
\vspace{-2mm}
\end{mathpar}
\caption{Selection of the clauses defining the unary interpretation and the relational interpretation of types.}
\label{fig:unary-interpret}
\end{figure*}

\paragraph{Relational Interpretation}
We give a selection of clauses for the definition of
the \emph{value interpretation} $\llp{\tau}{G,k}$ of relational types in
Figure~\ref{fig:unary-interpret}.  The interpretation of a relational
type is a set of pairs of related values. $G$, the Kripke world of the relational interpretation, is a mapping from
static array names $\gamma$ to 4-tuples $(l_1,l_2,n,\tau)$. If
$G(\gamma) =(l_1,l_2,n,\tau)$, then $l_1,l_2$ are the locations where
the arrays statically named $\gamma$ are stored in the two runs, $n$
is the length of these two arrays, and $\tau$ is the type at whose
relational interpretation the two arrays' corresponding elements
should be related. 
This is used for instance in the interpretation of the type
$\arr{\gamma}{n}{\tau}$. To define the interpretation of monadic types
we need a relation $(H_1,H_2)\vDash_{G,k} P $ which says that the
relational assertion $P$ holds of the heaps $H_1, H_2$ at world
$G$. We show the definition for the case
$P=\{\gamma \rightarrow \beta\}$ here:
\[
\begin{array}{r@{}l}
                 (H_1,H_2)\vDash_{G,k} (\gamma \rightarrow \beta)\ \ \mathsf{iff}\ \ & \exists l_1,l_2,\tau,n: G(\gamma)=(l_1,l_2,\tau, n) \\
& \land  \left( \forall i \leq n. \, \big (H_1(l_1)[i],H_2(l_2)[i]\big ) \in \llp{\tau }{G,k-1}\right)\\
& \land  \left(\forall i \leq n. \, \big (H_1(l_1)[i] \neq H_2(l_2)[i] \, \Rightarrow \, i \in \beta  \big) \right)\\
\end{array}
\]
Two heaps $H_1,H_2$ satisfy the assertion $\gamma \rightarrow \beta $
when all the related elements in the arrays $H_1(l_1)$ and $H_2(l_2)$
are in the value interpretation $\llp{\tau }{G,k-1}$ and,
additionally, for indices $i \not \in \beta$, the corresponding array
elements are \emph{equal}. Thus, $\beta$ tracks positions where the
two arrays \emph{may} differ, consistent with our earlier
description. Whether elements at indices in $\beta$ can actually
differ or not depends on $\tau$. For example, when $\tau$ is
$\tint[n]$ (for some $n$) or even $\exists i. \tint[i]$, this forces
corresponding elements to be equal at all indices (not just at those
outside $\beta$) since the relational interpretation of $\tint[n]$ is
the singleton $\{(n,n)\}$. However, when $\tau = U(A_1, A_2)$,
elements at indices not in $\beta$ can be arbitrary values of types
$A_1, A_2$ since the relational interpretation of $U(A_1, A_2)$ is
morally $A_1 \times A_2$.


The definition of the relational interpretation for a monadic type
$\monadR{P}{\exists \vec{\gamma}.\tau}{Q}{D}$
(Figure~\ref{fig:unary-interpret}) is the set of pairs of values
$(v_1,v_2)$ that when forced starting from heaps $H_1,H_2$ satisfying
the precondition $P$, result in heaps $H_1',H_2'$ satisfying the
postcondition $Q$. The relative cost of forcing must be upper-bounded
by $D$.
%
We extend the relational interpretation to pairs of expressions as follows:
$$
\begin{array}{l}
\llp{ \tau }{G,k}^{E,D} = \{  ( t_1, t_2 ) \, | \, \forall k_1 \leq k, v_1\, v_2.(\eval{t_1}{v_1}{c_1,k_1} \land \eval{t_2}{v_2}{c_2,k_2}) 
 \Rightarrow (v_1, v_2) \in \llp{\tau }{G,k-k_1} \land c_1-c_2 \leq D  \} 
\end{array}
$$
We also extend the interpretation to environments in the obvious way and prove a fundamental theorem for the relational typing. 
\setcounter{theorem}{1} \begin{theorem} [Fundamental Theorem for
  Relational Typing] 
If $\lenv;\senv;\cenv;\renv \cmp{t_1}{t_2}{\tau}{D}$ and $\vdash \delta: \Delta$ and  $\vDash \delta \Phi $ and $(\sigma_1,\sigma_2) \in \llp{\delta \Gamma}{G,k}$, then $ ( \delta \sigma_1 t_1,  \delta \sigma_2 t_1 ) \in \llp{\delta\tau}{G,k}^{E,(\delta D)} $. 
\end{theorem}


For readers familiar with Kripke logical relations, we note that our
worlds $g$ and $G$ are \emph{not} step-indexed (only our logical
relations are step-indexed). This is unlike some prior
work~\cite{DBLP:journals/jfp/NeisDR11,DBLP:conf/popl/TuronTABD13}. We
do not need step-indexed worlds since we include syntactic types, $A$
or $\tau$, for mutable locations (arrays) in the worlds. This suffices
for our purposes. Our logical relations are well-founded. The
relational interpretation for heaps, $(H_1,H_2)\vDash_{G,k} P $,
refers back to the value interpretation only at a step index strictly
smaller than $k$, while the value interpretation refers back to the
heap relation (via the clause for the monadic type,
$\monadu{P}{\exists{\vec{\gamma}}. A}{Q}{(L,U)}$) at a smaller or
equal step index. Consequently, the relation is well-founded by the
lexicographic order $\langle$step-index, size of type$\rangle$. Our
unary interpretation is well-founded for a similar reason.


\section{More Examples}
\label{sec:examples}
We discuss here three more examples demonstrating how we perform relational cost
analysis on programs with arrays. 
To improve readability, we omit some annotations and use syntactic sugar.

\paragraph*{Cooley Tukey FFT Algorithm}  This example shows how to use relational cost analysis to reason about \emph{constant-time} programs with imperative updates. We consider the following implementation of the Cooley Tukey algorithm for fast Fourier transforms~\cite{CooleyFFT}.  
$$
\begin{array}{l}
 \textsf{fix} \ \mathsf{FFT} \, (\_ ).  \lambda x.   \lambda y.  \lambda n. \lambda p.    \\
 \hspace{.2cm} \mathsf{if} \ 2 \leq n \ \mathsf{then} \  \mathsf{let} \ \{\_ \} = \mathsf{separate} \, ()  \, \,x \,n \,y \, p  \, \mathsf{in} \
  \mathsf{let} \ \{ \_ \} = \mathsf{FFT} \, ()\, x\, y\, (n/2)  \ p \ \mathsf{in} \\
 \hspace{0.8cm} \mathsf{let} \ \{ \_ \} = \mathsf{FFT} \, ()\, x\, y\, (n/2) \, (p+n/2)  \ \mathsf{in} \,  \mathsf{loop} () \,  0  \, n \,  x \, p  \ \mathsf{else} \  \mathsf{return} \, ()
 \end{array}
$$
$\mathsf{FFT}$ implements a divide-and-conquer algorithm. $x$ is the input array, $y$ is another array used for temporary storage, $n$ is the length of the two arrays, and $p$ is an index pointing to the index where the array should be split in the recursive call. This function uses a helper function $\mathsf{separate}$ to relocate elements in even positions to the lower half of the array $x$ and elements in odd positions to the upper half of the input array $x$ respectively, using $y$ as a scratchpad. We omit the code of $\mathsf{separate}$ here; it can be found in the Appendix. Another helper function $\mathsf{loop}$ simulates a for loop in which the input array $x$ is updated using the mathematical manipulations needed for the Fourier transform.
$$
\begin{array}{l}
\textsf{fix} \ \mathsf{loop} \, (\_ ). \lambda k.   \lambda n. \lambda x. \lambda p.  \  \mathsf{if} \ k< (n/2) \ \mathsf{then}  \\
 \hspace{.4cm}\mathsf{let} \, \{ e \} =\; \readx{x}{(k+p)} \ \mathsf{in} \ \mathsf{let} \, \{ o \} =\; \readx{x}{(k+p+n/2)} \, \mathsf{in} \\
 \hspace{.4cm}\mathsf{let} \, w = \mathsf{exp}(-2\pi k/n) \, \mathsf{in}\
\mathsf{let} \, \{ \_ \} =\; \updt{x}{(k+p)}{(e+w*o)} \, \mathsf{in}\, \\
 \hspace{.4cm}  \mathsf{let} \ \{ \_ \} =\; \updt{x}{(k+p+n/2)}{(e-w*o)} \ \mathsf{in} \ \mathsf{loop}\, () \, (k+1)  \, n \,  x \, p \
\mathsf{else}\,  \mathsf{return} \, ()
 \end{array}
$$
Intuitively, this example is constant time (for arrays with fixed length) because array manipulations depend only on the positions of the elements, and not on their values (assuming constant time addition and multiplication). 
One way to internalize this observation in the typing process is using only relational rules and relational types, always with relative cost $0$. We do this for the auxiliary functions $\mathsf{separate}$ and $\mathsf{loop}$ first, and with these we can easily give the following relational type to $\mathsf{FFT}$, witnessing the constant-time nature of this function.
\begin{multline*}
{  \jtypediff{0}{ \mathsf{FFT}}{\mathsf{FFT}}{} \begin{array}{l} \trunit \rightarrow
  \forall \gamma_1, \gamma_2, \beta_1 , M, N, P.  ( P+M < N  )  \supset  \\ \arr{\gamma_1}{N}{U(\tint, \tint) } \rightarrow   \arr{\gamma_2}{N}{U(\tint, \tint) } \rightarrow \tint[M]   \rightarrow\tint[P]  \rightarrow \\   \monadR{\gamma_1 \rightarrow \beta_1,\gamma_2 \rightarrow \mathbb{N} }{\exists \_. \trunit }{\gamma_1 \rightarrow \mathbb{N},\gamma_2 \rightarrow \mathbb{N}  }{  0 } 
\end{array}}
\end{multline*}
%
%
Another way to achieve the same result is to first compute the precise
lower and upper bounds on the unary cost of $\mathsf{FFT}$ and then
show that they are, in fact, equal.  However, computing the precise
unary cost of $\mathsf{FFT}$ is more difficult. We first establish the precise
cost of the helper functions. For example, we need to give the
following unary type to $\mathsf{loop}$.
\begin{multline*}
{  \vdash \mathsf{loop } : \begin{array}{l} \tunit \rightarrow
\forall \gamma_1, K,M, N, P.  ( P +M < N  )  \supset \tint[K]  \rightarrow \tint[M]  \rightarrow \arr{\gamma_1}{N}{\tint } \rightarrow \\  \tint[P] \rightarrow   \monadu{\gamma_1 \rightarrow \mathbb{N} }{\exists \_. \tunit }{\gamma_1 \rightarrow \mathbb{N} }{( 4*(M-K) , 4*(M-K)
)}
\end{array}}
\end{multline*}
Similarly, for the function $\mathsf{separate}$ we would need to
establish the precise cost, $4*M$. Once these unary costs are
available, we can conclude that the function $\mathsf{FFT}$ has the
same min and max costs: $8*M*\log(M)$ and is, thus, constant time
(using the rule \rname{R-S}).\footnote{It is not hard to see that
$\mathsf{FFT}$ has unary cost in $O(M*\log(M))$: The unary costs of
$\mathsf{separate}$ and the call to $\mathsf{loop()}$ are both linear
in $M$, so the cost $f(M)$ of $\mathsf{FFT}$ satisfies the recurrence
$f(M) = 2* f(M/2) + O(M)$, which has the standard solution
$O(M* \log(M))$. However, proving this in the type system is much
harder than the direct relational proof of $0$ relative cost.}

While both the unary and relational reasoning can show that this
example is constant time, the relational reasoning is \emph{much}
easier in this case since the relative cost is $0$ everywhere.

\paragraph*{Naive String Search} 
We show how a combination of unary and relational reasoning can give a precise relative cost to a class array algorithm: 
substring search.   
$$
\begin{array}{l}
\mathsf{NSS} = \textsf{fix} \ F \ (s). \  \lambda w. \lambda m. \lambda l_s. \lambda l_w. \lambda p. \\ 
 \hspace{0.4cm} \mathsf{if} \ (m+l_w)\leq l_s \ \mathsf{then}  \, \mathsf{let} \ \_ = \mathsf{search} \, s \, w \, m \, 0 \, l_s \, l_w \, p   \\ 
 \hspace{0.8cm} \mathsf{in} \,  F(s) \, w \, (m+1) \,  l_s \, l_w \,  p  \  \mathsf{else} \
 \mathsf{return} \, ()
\end{array}
$$
Here, strings are represented as arrays of integers (storing the ASCII code of each character). The function $\mathsf{NSS}$ takes as input, a ``long'' string \emph{s} and a ``short'' string \emph{w} in arrays, the lengths $l_s$ and $l_w$ of these arrays, and an array \emph{p} of length $l_s$ (we call this the result array). This function iteratively searches the substring \emph{w} at each position in \emph{s} and records whether the substring is found  at that position ($1$) or not ($0$).
To do this, $\mathsf{NSS}$ uses the following helper function $\mathsf{search}$.
$$
\begin{array}{l}
\textsf{fix} \ \mathsf{search} \, (\_ ). \, \lambda s. \,  \lambda w. \, \lambda m. \, \lambda i. \, \lambda l_s.\, \lambda l_w. \, \lambda p.  \\
\hspace{0.2cm} \mathsf{let} \, \{  x \} =\; \readx{s}{(m+i)} \, \mathsf{in} \
\mathsf{let} \,  \{ y \}=\; \readx{w}{i}  \, \mathsf{in} \\
\hspace{0.4cm} \mathsf{if} \, (i+1==l_w ) \, \mathsf{then}  \
 \big (\mathsf{if} \,   ( x==y ) \, \mathsf{then}  \, \updt{p}{m}{1}  \, \mathsf{else} \,   \updt{p}{m}{0}  \big ) \\
\hspace{0.4cm} \mathsf{else} \big( \mathsf{if} \, ( x==y ) \, \mathsf{then} \, \mathsf{ search} \, () \, s\, w \, m \, (i+1) \, l_s \, l_w \,  p\, \mathsf{else} \,  \updt{p}{m}{0} \big )
 \end{array}
$$
The function $\mathsf{search}$ has the same inputs as $\mathsf{NSS}$ except for the additional index $i$, that iterates over the positions of $l_w$. The two conditionals  check whether $\mathsf{search}$ is in its final step ($i+1 == l_w$), and whether the two corresponding characters in \emph{s} and \emph{w} coincide. When the two characters differ, \emph{p} is updated with $0$. When the two conditionals are satisfied at the same time,  \emph{p} is updated with $1$. 

Intuitively, $\mathsf{search}$ runs fastest when the first character of  \emph{w} does not appear in \emph{s}. It runs slowest when the suffix of \emph{w} starting at index $i$ occurs in \emph{s} at offset $m+i$. The difference of these two costs is a bound on the relative cost of $\mathsf{search}$. However, in the specific case where we consider two runs of $\mathsf{search}$ on the same string \emph{s}, the same index \emph{i} and where the two \emph{w}s agree on some prefix, we can see that the two runs behave identically until we reach an index where the \emph{w}s start to differ. In this case, we can give a better bound. To write this bound, we need to express the first index in the range $[i, l_w]$ where the two \emph{w}s differ. In {\THESYSTEM }, the index term $MIN(\beta_2 \cap [I, \infty ))$ represents this index (assuming $\beta_2$ is the relational pre-condition of \emph{w} and $I$ is the static index refinement for $i$'s size). Then, $\mathsf{search}$ incurs a nontrivial relative cost only \emph{after} this index is reached. Using this idea, we can show:
%
\begin{multline*}
{  \jtypediff{0}{ \mathsf{search}}{\mathsf{search}}{} \begin{array}{l} \tunit \rightarrow
\forall \gamma_1,\gamma_2,\gamma_3, I,M,R,N,\beta_2, \beta_3.  (I<R<N \land M+I<N )  \supset \\ \arr{\gamma_1}{N}{U(\tint) }  \rightarrow \arr{\gamma_2}{R}{U(\tint)} \rightarrow \tint[M] \rightarrow \tint[I]  \rightarrow \tint[N]  \Arrow{.2cm} \\ \tint[R]  \rightarrow  \arr{\gamma_3}{N}{U(\tbool)} \rightarrow   \monadR{P, \gamma_3 \Arrow{.2cm} \beta_3 }{\exists \_. \tunit}{P, \gamma_3 \Arrow{.2cm} \beta_3 \cup \{ M \} }{(R-1 -min(MIN(\beta_2 \cap [I, \infty ) ) , R-1 ) )*r  }
\end{array}}
\end{multline*}
%
where
$P= \gamma_1 \rightarrow \emptyset, \gamma_2 \rightarrow \beta_2$, $R$
is the static size of $l_w$, and $r$ is the (constant) cost of
two read operations. To account for the case where \emph{w} is the
same in the two executions we also add a lower bound $R-1$ in the
cost. The relative cost we establish here is more precise than the one
we would achieve with a non-relational analysis ($(R-1-I)*r$).

We stress here that to obtain this relative cost, the rule $\rname{R-FIX-EXT}$ is essential.  At a high level, typing proceeds by case analysis on $I\in\beta_2$. When  $I \not\in \beta_2$ we can proceed relationally with relative cost  $0$ in the recursive call. When $I \in \beta_2$ the control flows may differ in the two runs and we need to switch to unary reasoning via the rule $\rname{R-S}$. To obtain our bound using unary worst- best-case analysis  we need the precise unary type of $\mathsf{search}$, which is available in the context thanks to the rule $\rname{R-FIX-EXT}$.  The details of this proof are in our appendix.
%
By using the typing above for $\mathsf{search}$, we can also obtain an improved relative cost for $\mathsf{NSS}$ relative to itself: $(R-1 -min(MIN(\beta_2 \cap [I, \infty ) ) , R-1 ))*r*(N-M-R)$. This is simply the number of times $\mathsf{search}$ is called multiplied by the relative cost of $\mathsf{search}$.

\subsubsection*{Inplace Insertion Sort} Our next example, \emph{inplace insertion sort}, implements the insertion sort algorithm without any temporary arrays. The relative cost is complex but we can show that under reasonable assumptions, \THESYSTEM\ provides a more precise relative cost than a unary analysis. The algorithm is written in our language as follows.
\begin{align*}
 &\textsf{fix} \ \mathsf{ISort} \, (\_ ). \,   \lambda s. \,  \lambda i. \, \lambda l_s. 
\ \mathsf{if} \, (i<l_s) \, \mathsf{then} \, \mathsf{let} \ \{ a\} = \; \readx{s}{i} \ \mathsf{in}  \\
& \hspace{0.4cm}  \mathsf{let} \ \{b\} =  \mathsf{insert} \, () \,  s \, a \, 0 \, i \, \mathsf{in} \, \mathsf{ISort} \,() \, s \,  (i+1) \,  l_s  \  \mathsf{else} \ \mathsf{return} \, ()
 \end{align*} 
Intuitively, we observe that the cost of $\mathsf{ISort}$ relative to itself should be the sum of the possible cost variation of every recursive call, which is mainly decided by the auxiliary function $\mathsf{insert}$ below.   
\begin{align*}
& \textsf{fix} \ \mathsf{insert} \, (\_ ). \, \lambda s. \, \lambda a.\, \lambda x. \, \lambda i.   \ \mathsf{let} \, \{b\} = \; \readx{s}{x} \, \mathsf{in} \\
& \hspace{0.4cm} \mathsf{if} \, (a \geq b) \, \mathsf{then} \, \mathsf{insert} \, ()  \, s \, a \, (x+1) \, i    \\
&\hspace{0.4cm} \mathsf{else} \, 
\mathsf{let} \, \{ \_ \} = \mathsf{shift} \, ()  \, s \, x \, (i-1)  \, \mathsf{in}\,   \updt{s}{x}{a}  
 \end{align*}

The recursive function $\mathsf{insert}$ implements the standard operation of inserting an element into an array by finding the right position $x$ to insert the element at and shifting elements behind $x$ in the array backward before updating the value at index $x$ to $a$. This function uses a helper function $\mathsf{shift}$, which performs the shift operation. We omit the code here but note that $\mathsf{shift}$ uses one read operation and one update operation at every index, and finding the right position only needs one read operation. The unary cost of $\mathsf{ISort}$ is maximum when the input array is initially sorted in descending order. In contrast, the unary cost is minimum when the input array is initially sorted ascending. Assuming that read and update operations incur unit cost, the unary type of $\mathsf{insert}$ is as follows. 
\begin{multline*}
{  \vdash \mathsf{insert } : \begin{array}{l} \tunit \rightarrow
\forall \gamma_1. \forall N, X, I.  ( X \leq N \land I \leq N ) \supset\;  \arr{\gamma_1}{N}{\tint } \\  \rightarrow   \tint \rightarrow \tint[X] \rightarrow \tint[I] \rightarrow   \monadu{\gamma_1 \rightarrow \mathbb{N} }{\exists \_. \tunit}{\gamma_1 \rightarrow \mathbb{N} }{( I-X+1 ,2*(I-X)+2 )}
\end{array}}
\end{multline*}
%
%
With this unary type of $\mathsf{insert}$ in hand, we can obtain the relative cost of $\mathsf{insert}$ by switching to unary reasoning and then taking the difference. An interesting observation is that if the input arrays of the two runs coincide in the insertion range $[0,I]$ and the elements `$a$' being inserted also agree, then $\mathsf{insert}$'s cost relative to itself is $0$. The corresponding relational type is shown below.
\begin{multline*}
{  \jtypediff{0}{ \mathsf{insert}}{\mathsf{insert}}{} \begin{array}{l}  \trunit \rightarrow
  \forall \gamma_1, \beta_1. \forall N, A, X, I. ( X \leq N \land I \leq N \land \beta_1 \cap [X,I] = \emptyset )  \supset \\ \arr{\gamma_1}{N}{U (\tint) }   \rightarrow \trint \rightarrow  \tint[X] \rightarrow \tint[I]  \rightarrow   \monadR{\gamma_1 \Arrow{.2cm} \beta_1}{\exists \_. \trunit}{\gamma_1 \Arrow{.2cm} \beta_1  }{0  }
\end{array}}
\end{multline*}


This observation can be used in typing $\mathsf{ISort}$: For every
$I$, we split cases on whether $\beta_1 \cap [0,I] = \emptyset$ or not
(using rule \rname{R-P}). While $\beta_1 \cap [0,I] = \emptyset$, we
proceed relationally (with $0$ relative cost). Once $\beta_1 \cap
[0,I] \not= \emptyset$, we switch to unary reasoning using
rule \rname{R-S} since control flow may differ in the two runs. As in
the previous example, we need the rule \rname{R-FIX-EXT} for
this. Using this idea, we obtain a very precise relational cost for
$\mathsf{ISort}$.
\begin{multline*}
\jtypediff{0}{ \mathsf{ISort}}{\mathsf{ISort}}{} 
                                             \begin{array}{l}        \trunit \rightarrow
  \forall \gamma_1, \beta_1, N, I. ( I \leq N )  \supset \\ \arr{\gamma_1}{N}{U(\tint) }   \rightarrow \tint[I] \rightarrow \tint[N]  \rightarrow \monadR{\gamma_1 \rightarrow \beta_1 }{\exists \_. \trunit }{\gamma_1 \rightarrow \mathbb{N}   }{  \frac{ N*(N+1) - k*(k+1) }{2}  }
\end{array}
\end{multline*}
where the index term $k = max (I, min (MIN(\beta_1),N )) $ represents
the first index where the two arrays differ. The relative cost $\frac{
N*(N+1) - k*(k+1) }{2}$ is the sum of all the relative costs generated
in the recursive calls corresponding to indices in the range $[k,
N]$. Recursive calls up to index $k$ incur $0$ relative cost, as noted
above. More details are provided in the appendix. We note that the
cost obtained here is more precise than the relative cost that can be
obtained using unary reasoning alone.

\section{Bidirectional Type Checking}
\label{sec:btc}
Our next goal is to implement our type system {\THESYSTEM} to try out
our examples. However, implementing {\THESYSTEM} naively results in an
immediate challenge: Some of the rules are not syntax-directed and
lead to nondeterminism in an implementation. For example, in the split
rule \rname{R-P}, we can choose any constraint to split on (and this
is an infinite space); we can apply the switch rule \rname{R-S}
anywhere; in the rule \rname{R-FIX-EXT} we have to guess the unary
types of the functions (again an infinite space); and, there are two
rules for every array operator. To resolve this nondeterminism, we
introduce an extended expression language with annotations to guide
type-checking. For example, the term $(\esplit \, t \ewith C)$ marks
uses of rule \rname{R-P} that split on constraint $C$ in type-checking
$t$. The use of rule \rname{R-FIX-EXT} is indicated by the construct
$(\mathsf{FIXEXT} \,f(x). t \ewith \,\grt)$ that provides the
unary type $\grt$ of $(\mathsf{fix} f(x).t)$. Similarly, we introduce two variants of array
operations, e.g., $\mathsf{alloc}$ and $\mathsf{alloc_b}$
corresponding to the two rules \rname{R-L} and \rname{R-LB},
respectively.

Further, there is nondeterminism in subtyping due to the modal types
$\square$ and $U$. We resolve this by adding explicit type coercions
where subtyping would be needed. Prior work shows that this approach
is complete for these two modal types~\citep{birelcost}.

Beyond this, we face the usual challenge of algorithmizing any type
system: The need to either annotate or infer the types of bound
variables. Here, the problem is more nuanced than would be in a simply
typed or even a refinement type calculus, since we must also deal with
cost bounds in function and monadic types. To address this challenge,
we rely on the well-known middle ground of bidirectional type-checking
or local type inference~\cite{Pierce:2000}, where type annotations
must be provided only at explicit $\beta$-redexes and for top-level
functions, but everything else can be inferred. We design a
bidirectional type system for {\THESYSTEM} which is similar in spirit
to the one for \Relcost~\citep{birelcost}, but extended in nontrivial
ways to support bidirectional type-checking for array operations and
for our fixpoint extension.  This system can derive in an algorithmic
way four typing judgments, two relational and two unary.  The
relational typing judgment of
\THESYSTEM\ splits into two relational judgments in the bidirectional version, one for the ``checking mode'' and one
for ``inference mode''. The relational checking judgment has the form:
$$\lenv; \Delta; \psi_a; \Phi_a; \Gamma
\chdiff{t_1}{t_2}{\tau}{D}{\Phi}.$$
Given the location environment $\lenv$, the index variable environment
$\Delta$, the existential variable context $\Psi_a$, the current
constraint environment $\Phi_a$, the relational typing context $\Gamma
$, and terms $t_1$ and $t_2$, we \emph{check} against the relational
type $\tau$ and the relative cost $D$, and we generate the constraint
$\Phi$, which must be discharged separately. In contrast, the
relational inference judgment has the
form: $$\lenv; \Delta; \psi_a; \Phi_a; \Gamma \infdiff{t_1}{t_2}{\tau}{\psi}{D}{\Phi}.$$
Here, we \emph{synthesize} the relational type $\tau$ and the relative
cost $D$, and we generate the constraint $\Phi$ with all the newly
generated (existential) variables in $\psi$. We have similar judgments
for the unary case. The unary checking judgment has the form
$\lenv;\Delta; \psi_a; \Phi_a; \Omega \chexec{t}{\grt}{L}{U}{\Phi}$,
while the unary inference judgment has the form
$\lenv;\Delta; \psi_a; \Phi_a; \Omega \infexec{t}{\grt}{\psi}{L}{U}{\Phi}$. Both
these judgments can be understood in a way similar to their relational
counterparts. In all the judgments, we write all the outputs (inferred
components) in \textcolor{red}{red} and inputs in black. Notice that
in comparison with the typing judgments in \THESYSTEM, the algorithmic
typing judgments have one more input context $\Psi_a$ which records
previously eliminated existential variables.

We show selected algorithmic typing judgments in
Figure~\ref{fig:algo-rule-pure} to explain how we handle
{\THESYSTEM}'s non-determinism. The switch rule (\rname{R-S}) exists in
both checking and inference modes because we find it convenient to use
the rule in both modes in our examples. Both algorithmic rules relate
the annotated terms $(\eswitch t_1)$ and $(\eswitch t_2)$ at the type
$\tcho{(\grt_1, \grt_2)}$ and generate the final constraint based on
the constraints from subterms $t_1$ and $t_2$ obtained in unary mode.
The relative cost $D$ must be the difference of the maximal unary cost
of $t_1$ ($U_1$) and the minimal unary cost of $t_2$ ($L_2$). In the
checking rule, \textbf{alg-r-switch$\downarrow$}, this is forced in
the output constraint. The split rule (\rname{R-P}) exists only in
checking mode (\textbf{alg-r-split$\downarrow$}). The terms
$(\esplit \, t_1 \ewith C)$ and $(\esplit \, t_2 \ewith C)$ determine
that this rule must be applied, splitting on constraint $C$. The final
output constraint $\cand{\cimpl{C}{\Phi_1}}{\cimpl{\neg C}{\Phi_2}}$
also analyzes $C$.
The algorithmic counterpart of the
rule \rname{R-FIX-EXT} in checking mode, \textbf{alg-fixext$\downarrow$}, relates the annotated terms $(\mathsf{FIXEXT} \,f(x). t
\ewith \,\grt_1)$ and $(\mathsf{FIXEXT} \,f(x). t \ewith \,\grt_1)$
and checks the subterms $\mathsf{fix} f(x).t$ and $\mathsf{fix}
f(x).t'$ at the unary types $\grt_1$ and $\grt_2$, respectively. The final constraint is the combination
of the constraints generated from the unary checking of the two
subterms and the relational checking of the two function bodies.

\begin{figure*}
\begin{mathpar}
\inferrule
  {
    \Delta; \psi_a; \Phi_a; \trmo{\Gamma}  \infexec{t_1}{\grt_1}{\psi_1}{\_}{U_1}{\Phi_1}\\
    \Delta; \psi_a; \Phi_a; \trmo{\Gamma}  \infexec{t_2}{\grt_2}{\psi_2}{L_2}{\_}{\Phi_2}
  }
  {
    \bctx  \infdiff{\eswitch t_1}{\eswitch t_2}{\tcho{(\grt_1, \grt_2)}}{\psi_1,\psi_2}{U_1-L_2}{\cand{\Phi_1}{\Phi_2
  }}}~\textbf{alg-r-switch$\uparrow$}
\and
\inferrule
  {
    \freshCost{L_1, U_1, L_2, U_2} \\
    \Delta; U_1, L_1, \psi_a; \Phi_a; \trmo{\Gamma} \chexec{t_1}{\grt_1}{L_1}{U_1}{\Phi_1}\\
    \Delta; U_2, L_2, \psi_a; \Phi_a; \trmo{\Gamma}
    \chexec{t_2}{\grt_2}{L_2}{U_2}{\Phi_2} \quad
    \Phi = \cand{\Phi_1}{\cexists{L_2,U_2}{\scost}{\cand{\Phi_2}{\ceq{U_1-L_2}{D}}} }
  }
  {
    \lenv ; \bctx  \chdiff{\eswitch t_1}{\eswitch t_2}{\tcho{(\grt_1, \grt_2)}}{D}{\cexists{L_1,U_1}{\scost}{(\Phi)}}
   }~\textbf{alg-r-switch$\downarrow$}
\and
\inferrule{
  \lenv;\Delta; \psi_a; C \wedge \Phi_a; \Gamma \chdiff{t_1}{t_1'}{\tau}{D}{\Phi_1}\\
  \lenv;\Delta; \psi_a; \neg C \wedge \Phi_a; \Gamma \chdiff{t_1}{t_1'}{\tau}{D}{\Phi_2}\\
  \Delta  \wfcs{C}
}
{ 
  \lenv;\bctx \chdiff{\esplit (t_1) \ewith C}{\esplit (t_1') \ewith C}{\tau}{D}{\cand{\cimpl{C}{\Phi_1}}{\cimpl{\neg C}{\Phi_2}}}
} ~\textbf{alg-r-split$\downarrow$}
\end{mathpar}
\begin{flushleft} \textbf{alg-fixext$\downarrow$} \end{flushleft}
\begin{mathpar}
\inferrule
{
{\Delta; \psi_a; \Phi_a; \trmo{\Gamma}
  \chexec{\mathsf{fix} f(x). t}{\grt_1}{0}{0}{\Phi_1}  } \\
{\Delta; \psi_a; \Phi_a; \trmo{\Gamma} \chexec{\mathsf{fix} f(x).t'}{\grt_2}{0}{0}{\Phi_2}  } \\
 \lenv;\Delta; {\psi_a};  \Phi_a; f : \tau_1 \tarrd{D'} \tau_2,f:
 U(\grt_1,\grt_2),  x:\tau_1, \Gamma \chdiff{t}{t'}{\tau_2}{D'}{\Phi}\\
\Phi_r = \Phi \land \Phi_1 \land \Phi_2
}
{
\lenv; \bctx \chdiff{\mathsf{FIXEXT} f(x). t \ewith  \grt_1 }{\mathsf{FIXEXT} f(x). t' \ewith  \grt_2 }{\tau_1 \tarrd{D'} \tau_2}{D}{\cand{\Phi_r}{\ceq{0}{D}}}
} 
%
%
\end{mathpar}
\caption{Selection of algorithmic typing rules}
\label{fig:algo-rule-pure}
\end{figure*}

\begin{figure*}

\begin{flushleft}~\textbf{alg-r-alc-$\downarrow$}  \end{flushleft} 
\begin{mathpar}
 \inferrule
{
 \freshCost{D_1,D_2}
 \\
 \lenv;\Delta; {D_1,\psi_a};  \Phi_a;  \Gamma \chdiff{t_1}{t_1'}{\tint[I]}{D_1}{\Phi_1}
 \\
 \lenv;\Delta; {D_2,\psi_a};  \Phi_a;  \Gamma \chdiff{t_2}{t_2'}{\tau }{D_2}{\Phi_2}
 \\
 \Phi =  \Phi_2 \land \ceq{D_1+D_2}{D}
 \\
    {\lenv \vdash \gamma\ \mathsf{fresh} }
    \\
    {\lenv ;\senv  \wf{P} } \\
    \Phi_r = \cexists{D_1}{\scost}{ \Phi_1 \land (\cexists{D_2}{\scost}{\Phi} ) }
}
{
\lenv;\bctx \chdiff{\ealloc{t_1}{t_2} }{\ealloc{t_1'}{t_2'} }{
\mathrel{  \monadR{P}{\exists \gamma.\arrR{\gamma}{I}{\tau}}{ P \star \gamma \rightarrow \mathbb{N} }{D}  }
}{0}{ \Phi_r
  }
} 
\end{mathpar} 
\begin{flushleft}~\textbf{alg-r-alcB-$\downarrow$}  \end{flushleft} 
\begin{mathpar}
\inferrule
{
 \freshCost{D_1,D_2}
 \\
 \lenv;\Delta; {D_1,\psi_a};  \Phi_a;  \Gamma \chdiff{t_1}{t_1'}{\tint[I]}{D_1}{\Phi_1}
 \\
 \lenv;\Delta; {D_2,\psi_a};  \Phi_a;  \Gamma \chdiff{t_2}{t_2'}{\tbox {\tau} }{D_2}{\Phi_2}
 \\
 \Phi =  \Phi_2 \land \ceq{D_1+D_2}{D}
 \\
    {\lenv \vdash \gamma\ \mathsf{fresh} }
    \\
    {\lenv ;\senv  \wf{P} } 
    \\
    \Phi_r = \cexists{D_1}{\scost}{ \Phi_1 \land (\cexists{D_2}{\scost}{\Phi} ) }
}
{
\lenv;\bctx \chdiff{\eallocB{t_1}{t_2} }{\eallocB{t_1'}{t_2'} }{
\mathrel{  \monadR{P}{\exists \gamma.\arrR{\gamma}{I}{\tau}}{ P \star \gamma \rightarrow \emptyset }{D}  }
}{0}{ \Phi_r
  }
} 
\end{mathpar} 

\begin{flushleft}~\textbf{alg-r-read-$\downarrow$}  \end{flushleft} 
\begin{mathpar}
\inferrule
{
 \Delta; {\psi_a};  \Phi_a;  \Gamma \infdiff{t_1}{t_1'}{\arrR{\gamma}{I}{\tau} }{\psi_1}{D_1}{\Phi_1}
 \\
 \Delta; {\psi_1,\psi_a};  \Phi_a;  \Gamma \infdiff{t_2}{t_2'}{\tint[I'] }{\psi_2}{D_2}{\Phi_2}
 \\
P = P'\star \gamma \rightarrow \_ 
\\
 \Delta; {\psi_a};  \Phi_a \vDash  \cleq{I'}{I} 
 \\
 \Phi =  \Phi_2 \land \ceq{D_1+D_2}{D}
    \\
    {\lenv ;\senv  \wf{P} } \\
    \Phi_r = \exists (\psi_1). \Phi_1 \land (  \exists (\psi_2).\Phi )
}
{
\bctx \chdiff{\ereadx{t_1}{t_2} }{\ereadx{t_1'}{t_2'} }{
\mathrel{  \monadR{P}{\exists \_.\tau }{ P }{D}  }
}{0}{
{\Phi_r }   }
} 
\end{mathpar}
\begin{flushleft}~\textbf{alg-r-readB-$\downarrow$}  \end{flushleft} 
\begin{mathpar}
\inferrule
{
 \Delta; {\psi_a};  \Phi_a;  \Gamma \infdiff{t_1}{t_1'}{\arrR{\gamma}{I}{\tau} }{\psi_1}{D_1}{\Phi_1}
\\
P = P'\star \gamma \rightarrow \beta 
\\
 \Delta; {\psi_1,\psi_a};  \Phi_a;  \Gamma \infdiff{t_2}{t_2'}{\tint[I'] }{\psi_2}{D_2}{\Phi_2}
\\
 \Delta; {\psi_a};  \Phi_a \vDash  \cleq{I'}{I}  \land \neg( I' \in \beta)
 \\
 \Phi =  \Phi_2  \land  \ceq{D_1+D_2}{D}  
    \\
    {\lenv ;\senv  \wf{P} } \\
    \Phi_r =\exists (\psi_1). \Phi_1 \land (  \exists (\psi_2).\Phi )
}
{
\bctx \chdiff{\ereadxB{t_1}{t_2} }{\ereadxB{t_1'}{t_2'} }{
\mathrel{  \monadR{P}{\exists \_.\square \tau }{ P }{D}  }
}{0}{
{ \Phi_r}   }
} 
\end{mathpar}
\begin{flushleft}~\textbf{alg-r-updt-$\downarrow$}  \end{flushleft} 
\begin{mathpar}
\inferrule
{
 \Delta; {\psi_a};  \Phi_a;  \Gamma \infdiff{t_1}{t_1'}{\arrR{\gamma}{I}{\tau} }{\psi_1}{D_1}{\Phi_1}
 \\
 \freshCost{D_3}
\\
 \Delta; {\psi_1,\psi_a};  \Phi_a;  \Gamma \infdiff{t_2}{t_2'}{\tint[I'] }{\psi_2}{D_2}{\Phi_2}
 \\
 \Delta;\psi_a; \Phi_a  \vDash  {\cleq{I' }{I}} \land \beta' = \beta \cup \{I'\}
 \\
 \Delta; {D_3,\psi_2,\psi_1,\psi_a};  \Phi_a;  \Gamma \chdiff{t_3}{t_3'}{ \tau  }{D_3}{\Phi_3}
 \\
P = P'\star \gamma \rightarrow \beta
 \\
 Q = P'\star \gamma \rightarrow \beta'
 \\
 \Phi =  \Phi_2 \land \ceq{D_1+D_2+D_3}{D} 
    \\
    {\lenv ;\senv  \wf{P'} } \\
    \Phi_r = \exists (\psi_1).(
\Phi_1 
\land ( \exists (\psi_2). (
\Phi_2 \land \cexists{D_3}{\scost}{\Phi}
) )
}
{
\bctx \chdiff{\eupdt{t_1}{t_2}{t_3} }{\eupdt{t_1'}{t_2'}{t_3'} }{
\mathrel{  \monadR{P}{\exists \_.\tunit }{ Q }{D}  }
}{0}{ \Phi_r
   }
} 
\end{mathpar}
\begin{flushleft}~\textbf{alg-r-updtB-$\downarrow$}  \end{flushleft} 
\begin{mathpar}
 \inferrule
{
 \Delta; {\psi_a};  \Phi_a;  \Gamma
 \infdiff{t_1}{t_1'}{\arrR{\gamma}{I}{\tau} }{\psi_1}{D_1}{\Phi_1}
\\
 \freshCost{D_3}
 \\
 \Delta; {\psi_1,\psi_a};  \Phi_a;  \Gamma \infdiff{t_2}{t_2'}{\tint[I'] }{\psi_2}{D_2}{\Phi_2}
  \\
 \Delta;\psi_a; \Phi_a  \vDash  {\cleq{I' }{I}} \land \beta' = \beta \setminus \{I'\}
 \\
 \Delta; {D_3,\psi_2,\psi_1,\psi_a};  \Phi_a;  \Gamma \chdiff{t_3}{t_3'}{ \tbox{ \tau}  }{D_3}{\Phi_3}
 \\
P = P'\star \gamma \rightarrow \beta
 \\
 Q = P'\star \gamma \rightarrow \beta'
 \\
 \Phi =  \Phi_2 \land \ceq{D_1+D_2+D_3}{D} 
    \\
    {\lenv ;\senv  \wf{P'} } \\
    \Phi_r = \exists (\psi_1).(
\Phi_1 
\land ( \exists (\psi_2). (
\Phi_2 \land \cexists{D_3}{\scost}{\Phi}
) ) 
}
{
\bctx \chdiff{\eupdtB{t_1}{t_2}{t_3} }{\eupdtB{t_1'}{t_2'}{t_3'} }{
\mathrel{  \monadR{P}{\exists \_.\tunit }{ Q }{D}  }
}{0}{ \Phi_r
  }
} 
\end{mathpar}
\caption{Selection of algorithmic typing rules for array operations}
\label{fig:algo-rule}
\end{figure*}


%
%

Next, we discuss selected rules for array operations. These operations
constitute the main challenge in our bidirectional type system
relative to prior work. We show a selection of bidirectional rules for
array operations in Figure~\ref{fig:algo-rule}. As mentioned, to
resolve the non-determinism between the $\square$-ed and non-$\square$-ed
rules for each array operation, we use distinct expressions, e.g.,
$\eallocB{t_1}{t_2}$ vs $\ealloc{t_1}{t_2}$. To start understanding
the rules, note that the conclusion of every array operation is typed
in checking mode. The two allocation
rules \textbf{alg-r-alc-$\downarrow$}
and \textbf{alg-r-alcB-$\downarrow$} check the first arguments $t_1$
and $t_1'$ against the relational type $\tint[I]$ and relative cost
$D_1$, then check the second arguments $t_2$ and $t_2'$ against the
relational type $\tau$ (or $\square \tau$) and relative cost
$D_2$. The final constraint $\Phi_r
= \cexists{D_1}{\scost}{ \Phi_1 \land (\cexists{D_2}{\scost}{\Phi} ) }
$ requires that there exist $D_1$ and $D_2$ such that $\Phi_1$ and
$\Phi_2$ hold and that $D_1+D_2$ equals the given cost $D$.  The
algorithmic typing rules for $\mathsf{read}$ and $\mathsf{updt}$ have
other interesting aspects. These rules are in checking mode but the
types of the first two arguments are inferred, not checked. This is
because, although we know that the first argument of
$\ereadx{t_1}{t_2}$ or $\eupdtB{t_1}{t_2}{t_3}$ must be an array and
the second argument must be a number, we do not know the size of the
array or the size (refinement index) of the number. Hence, we must
infer this information. Additionally, these rules make checks on the
pre- and post-conditions. As an example, the condition
$ \neg(I' \in \beta) $ is checked on the
rule \textbf{alg-r-readB-$\downarrow$} to guarantee that we are indeed
reading the same element on the two sides. Similarly, in the
rules \textbf{alg-r-updt-$\downarrow$}
and \textbf{alg-r-updtB-$\downarrow$}, the $\beta'$ in the
post-condition, representing the differences between the two arrays,
must be the same as the $\beta$ in the pre-condition except for the
index $I'$ which has been updated. For this, in the
rule \textbf{alg-r-updt-$\downarrow$} we check that $\beta'
= \beta \cup \{I'\}$, while in the
rule \textbf{alg-r-updtB-$\downarrow$} we check that $\beta'
= \beta \setminus \{I'\}$, consistent with the (non-algorithmic)
typing rules.

\section{Implementation and Experiments}
\label{sec:exp}
We implemented the bidirectional type checking system for {\THESYSTEM}
described in Section~\ref{sec:btc}. Using this implementation, we
checked all the examples described in this paper as well as some
others that are described in the appendix. We explain the results of
our experiments in this section. One small difference between the
system described in Section~\ref{sec:btc} and our implementation is
that rather than support two syntactic variants of every array
operation, we use heuristics to infer whether to apply the
$\square$-ed rule or the non-$\square$-ed rule. For example, to decide
to apply the rule \textbf{alg-r-readB-$\downarrow$} as opposed
to \textbf{alg-r-read-$\downarrow$}, we check that
$I \not\in \beta$. We always try the $\square$-ed rules first. These
heuristics suffice for our examples and reduce our annotation burden
at the cost of some extra constraint solving time. Our typechecker is
implemented in OCaml and we plan to open source it.

\paragraph{Constraint Solving}
The primary difficulty in our implementation and the most
time-consuming step in type checking is solving the constraints that
the bidirectional type system generates. For this we rely on an SMT
solver. Specifically, we use Alt-Ergo~\citep{bobot2013alt} through the
Why3 frontend~\citep{Filliatre:2013}. A fundamental difficulty here is
that the SMT solver struggles with constraints that have too many
existential quantifiers. To alleviate this concern, we rely on a
solution proposed in the implementation of \Relcost~\citep{birelcost}:
We implement a simple algorithm that generates candidate substitutions
for existentially quantified variables by examining equality and
inequality constraints that mention the variables. This works
remarkably well (we refer to~\citep{birelcost} for details). A new
challenge for \THESYSTEM\ is how to represent and solve constraints
involving the sets $\beta$. For this, we rely on the library for set
theory from Why3.


\paragraph{Experiments}
Table~\ref{tab:rel} summarizes some statistics about the performance
of our type checker on different examples.
\rd{For each example, we show the number of lines of code (LOC), the
number of type annotations that are needed (\#TYP), the number of annotations needed to disambiguate rules (\#ESF), the time needed for
type checking (TC), the time needed for solving the constraints that
arise as premises during type checking (TC-SMT), and the time needed
for solving the final constraint which is the output of the type
checking (TF-SMT).} Our experiments were performed on a 3.1 GHz Intel
Core i5 processor with 8GB of RAM.

The programs
$\mathsf{map(1)}$, $\mathsf{map(2)}$, $\mathsf{boolOr}$,
$\mathsf{FFT}$, $\mathsf{NSS}$ and $\mathsf{ISort}$ are implementations of the
corresponding examples discussed in Section~\ref{sec:overview} and 
Section~\ref{sec:examples}. For $\mathsf{FFT}$, which uses the auxiliary
functions $\mathsf{separate}$ and $\mathsf{loop}$, we report 
statistics for the whole program and individually
for each auxiliary function. 
The program $\mathsf{ISort}$ uses helper functions $\mathsf{insert}$
and $\mathsf{shift}$. These are also shown separately. The programs
$\mathsf{merge(1)}$ and $\mathsf{merge(2)}$ consider an imperative
versions of merge sort, typed with two different relational types.
The function $\mathsf{SAM}$ (square-and-multiply) computes a positive
power of a number represented as an array of bits, while
$\mathsf{comp}$ checks the equality of two passwords represented as
arrays of bits. These last two examples are array-based
implementations of similar list-based implementations presented
in~\citet{Cicek17}. More details are in the
appendix.

\begin{table}[h]
    \caption{Summary of experimental results}
    \label{tab:rel}
    \centering
    \begin{tabular}{| p{2cm} | p{1cm} | p{1cm} | p{1cm} |
      p{1.5cm} | p {1.5cm}| p {1.5cm}|}
\hline
      Benchmark  &  LOC &\#TYP & \#ESF & TC & TC-SMT & TF-SMT 
                                                                            \\
      \hline
   $\mathsf{map(1)}$ &  19  & 3  & 0 &    0.802s  & 1.051s & 0.01s   \\
    $\mathsf{map(2)}$ &  12  & 2 & 1 &    1.247s  & 0.994s & 0.02s  \\ \hline
 $\mathsf{boolOr}$ &  48  & 8 & 3 &    1.574s  & 1.131 & 2.38s \\  \hline
 $\mathsf{separate} $& 36   &  8 & 0 & 1.351s & 2.148s & 0.01s \\ 

     $\mathsf{loop}$ & 23  & 5 & 0  & 1.167s & 2.114s & 0.01s \\ 
     $\mathsf{FFT}$   & 66   & 17  & 0 &  2.591s & 4.268s & 0.01s \\ \hline
     $\mathsf{Search} $ &  62  & 10 & 3 &   3.753s  & 4.43s & 6.56s \\
    $\mathsf{NSS} $ &  94  & 12 & 3 &   4.158s  & 4.413s & 10.03s \\  \hline
    $\mathsf{shift}$ &  14  & 3 & 0 &    0.660s & 1.394s &  0.01s \\
    $\mathsf{insert}$ & 22 & 6 & 0 & 1.001s & 3.019s & 0.01s \\  
    
 $\mathsf{iSort}$ & 134 & 12 & 3 & 2.897s  & 6.181s & 10.70s \\  \hline
     $\mathsf{merge(1)}$ &  29  & 8 & 0  &    2.203s  & 2.232s & 0.01s  \\
      $\mathsf{merge(2)} $ &  64  & 11 & 2 &   3.231s & 0.349s & 0.02s \\ \hline

   
    $\mathsf{sam}$ &  19  & 4 & 1 &    0.946s  & 0.083s & 0.02s \\ \hline
    $\mathsf{comp}$ &  20  & 3  & 0 &    1.138s  & 0.112s & 0.01s \\ \hline
    \end{tabular}
\end{table}
The results in Table~\ref{tab:rel} show that {\THESYSTEM} can be used
effectively to reason about the relative cost of functional-imperative
programs. Unsurprisingly, examples combining relational and unary
reasoning (using rules \rname{R-FIX-EXT} and
\rname{R-S}) such as $\mathsf{boolOr}$, $\mathsf{NSS}$
and $\mathsf{ISort}$ need more annotations and need more time for both
type checking and SMT solving. 
In some examples like $\mathsf{ISort}$,
TC-SMT, the time taken for solving constraints in the premises of the
rules is very high. This is because of the heuristic we described at
the beginning of this section where we try $\square$-ed rules before
non-$\square$-ed rules. The SMT solver first tries to prove that the
$\square$-ed rule can be applied, but in some cases it \emph{times
out}. This timeout period is counted in TC-SMT. It is set to 1s in all
examples, except $\mathsf{ISort}$ and $\mathsf{Insert}$, where we try
for 2s. TF-SMT, the time taken to check the final output constraint,
is also high for some examples like $\mathsf{ISort}$, but this is due
to the complexity of the constraint. Further improving our heuristics
and the constraint solving process remains a direction for future
work.

\section{Related work}

A lot of prior work has studied static cost analysis. We discuss some
of this work here. \citet{Reistad:1994} present a type and effect
system for cost analysis where, like {\THESYSTEM}, the cost can
depend on the size of the input.
\citet{Danielsson08} uses a cost-annotated monad similar in spirit to
the one we use here.  \citet{lago12} present a linear dependent type system
using index terms to analyze time complexity. 
\citet{HoffmannAH12} present an automated amortized cost
analysis for programs with complex data structures such as matrices.
\citet{wang17:timl} develop a type system for cost analysis with
time complexity annotations in types. However, none of these systems
consider relational analysis of costs.

\citet{ChargueraudP15} present an amortized resource
analysis based on an extension of separation logic with time
credits. Our use of triples and separation-based management of arrays
references is similar to theirs. However, their
technique is based on separation logic, while ours is based
on a type-and-effect system. Moreover, they consider only unary reasoning
while we are interested primarily in relational reasoning.
\citet{DBLP:conf/rta/LichtmanH17} present an amortized
resource analysis for arrays and reference based on arrays with
potentials. Their technique represents the available ``potential''
before and after a computation, similar to our triples. Again, they
focus only on unary cost analysis and, consider mostly
first-order programs and linear potentials.

Outside of cost analysis, a lot of work has considered relational
verification techniques for other applications. \citet{LahiriVH10}
present a differential static analysis to find code defects looking at
two pieces of code relationally.  Probabilistic relational
verification has seen many applications in
cryptography~\citep{BFGSSZ14} and differential
privacy~\citep{gaboardi13:dep,BartheGAHRS15}. The indexed types used
by
\citet{gaboardi13:dep} are similar in spirit to ours. 
\citet{DBLP:conf/asplos/ZhangWSM15} introduce dependent labels into the
type of SecVerilog, an extension of Verilog with information flow
control. The use of a lightweight invariant on variables and security
levels in SecVerilog is similar to our use of $\beta$, which is also
an invariant on static location
variables. \citet{DBLP:conf/cav/UnnoTS17} present an automated
approach to verification based on induction for Horn clauses, which can also
be used for relational verification.
Benton et al.~\cite{DBLP:conf/popl/Benton0N14,DBLP:conf/ppdp/Benton0N16} introduce  abstract
effects to reason about abstract
locations.
This is conceptually similar to the way our preconditions and
postconditions allow us to reason about different independent
locations.

Our work is inspired by \Relcost~\citep{Cicek17} and
DuCostIt~\citep{DuCostIt16}. These are refinement type and effect
systems for pure functional languages \emph{without mutable
state}. \Relcost\ supports relational cost analysis of pure
programs. In contrast, \THESYSTEM\ supports imperative arrays. The
difference is substantial: Besides significant changes to the model,
the type system has to be enriched with Hoare-like triples, whose
design is a key contribution of our work. \Relcost\ has an
implementation via an SMT back-end~\cite{birelcost}; we extend this
approach with imperative features and support for sets of indices (our
$\beta$s).
\citet{NDFH17} combine information flow and
amortized resource analysis to guarantee constant-resource
implementations. Their type system allows relational reasoning about resources
through precise unary analysis.  Their focus is on first-order
functional programs and on the constant time guarantee, while we want
to support functional-imperative programs and more general relative
costs.
\citet{DBLP:journals/pacmpl/RadicekBG0Z18} add a
cost monad to a relational refinement type system, where refinements
reason about relational cost, for programs without
state. This system is expressive: it supports a combination of
cost analysis with value-sensitivity and full functional
specifications (\Relcost\ can also be embedded in it). However, it
requires a framework for full functional verification.  Our approach
is complementary in that we use lighter refinements that are easier to
implement, but do not support full functional verification.

\section{Conclusion}
We presented \THESYSTEM, a relational type-and-effect system that can
be used to reason about the relative cost of functional-imperative
programs with mutable arrays. Our key contribution is a set of
lightweight relational refinements allowing one to establish different
relations between pairs of state-affecting computations, including
upper-bounds on cost difference. We have discussed how {\THESYSTEM} is
implemented and used {\THESYSTEM} to reason about the relational cost
of several nontrivial examples.


\begin{acks}                            
  This work is in part supported by the
  \grantsponsor{GS100000001}{National Science
    Foundation}{http://dx.doi.org/10.13039/100000001} under Grant
  No.~\grantnum{GS100000001}{1718220}. 
\end{acks}

\bibliography{main}




\end{document}